\documentclass[
twocolumn,
superscriptaddress,
amsmath,amssymb,
prl,
]{revtex4-2}
\usepackage{graphicx}
\usepackage{hyperref}
\usepackage{dcolumn}
\usepackage{bm}
\usepackage{epstopdf}
\usepackage{romannum}
\usepackage{csquotes}
\usepackage[dvipsnames]{xcolor}
\usepackage{hyperref}
\usepackage{float}

\begin{document}
	
\preprint{}
	
\title{Bulk and surface Dirac states accompanied by two superconducting domes in FeSe-based superconductors}
	
\author{Qiang Hou}
\affiliation{Key Laboratory of Quantum Materials and Devices of Ministry of Education, School of Physics, Southeast University, Nanjing 211189, China}
\author{Wei Wei}
\affiliation{Key Laboratory of Quantum Materials and Devices of Ministry of Education, School of Physics, Southeast University, Nanjing 211189, China}
\author{Xin Zhou}
\affiliation{Key Laboratory of Quantum Materials and Devices of Ministry of Education, School of Physics, Southeast University, Nanjing 211189, China}
\author{Wenhui Liu}
\affiliation{Key Laboratory of Quantum Materials and Devices of Ministry of Education, School of Physics, Southeast University, Nanjing 211189, China}
\author{Ke Wang}
\affiliation{Key Laboratory of Quantum Materials and Devices of Ministry of Education, School of Physics, Southeast University, Nanjing 211189, China}
\author{Xiangzhuo Xing}
\affiliation{School of Physics and Physical Engineering, Qufu Normal University, Qufu 273165, China}
\author{Yufeng Zhang}
\affiliation{School of Physics and Electronic Engineering, Jiangsu University, Zhenjiang 212013, China}
\author{Nan Zhou}
\affiliation{Key Laboratory of Materials Physics, Institute of Solid State Physics, HFIPS, Chinese Academy of Sciences, Hefei, 230031, China}
\author{Yongqiang Pan}
\affiliation{Key Laboratory of Materials Physics, Institute of Solid State Physics, HFIPS, Chinese Academy of Sciences, Hefei, 230031, China}
\author{Yue Sun}
\email{Corresponding author: sunyue@seu.edu.cn}
\affiliation{Key Laboratory of Quantum Materials and Devices of Ministry of Education, School of Physics, Southeast University, Nanjing 211189, China}
	
\author{Zhixiang Shi}
\email{Corresponding author: zxshi@seu.edu.cn}
\affiliation{Key Laboratory of Quantum Materials and Devices of Ministry of Education, School of Physics, Southeast University, Nanjing 211189, China}

\begin{abstract}

\textbf{}
Recent investigations of FeSe-based superconductors have revealed the presence of two superconducting domes, and suggest possible distinct pairing mechanisms. Two superconducting domes are commonly found in unconventional superconductors and exhibit unique normal states and electronic structures. In this study, we conducted electromagnetic transport measurements to establish a complete phase diagram, successfully observing the two superconducting domes in FeSe$_{1-x}$S$_x$ (0 $\le x \le$ 0.25) and FeSe$_{1-x}$Te$_x$ (0 $\le x \le$ 1) superconductors. The normal state resistivity on SC1 shows the strange metal state, with a power exponent approximately equal to 1 ($\rho (T)\propto T^n$ with $n\sim 1$), whereas the exponent on SC2 is less than 1. A bulk Dirac state observed on SC1, completely synchronized with the strange metal behavior, indicating a close relationship between them. While a topological surface Dirac state is witnessed on SC2, and undergoes a sign change near the pure nematic quantum critical point. The evolution of the Dirac states indicates that the appearance of the two superconducting domes may originate from the Fermi surface reconstruction. Our findings highlight distinct Dirac states and normal state resistivity across the two superconducting domes, providing convincing evidence for the existence of the two different pairing mechanisms in FeSe-based superconductors.
\end{abstract}
	
	
\maketitle

\textbf{}
Two superconducting domes are commonly found in unconventional superconductors such as cuprates \cite{twodomeCu,twodomeCu214,twodomeCu2141992}, iron-based compounds \cite{twodomesun,twodomeiron2014} and heavy fermions \cite{twodomeheavy}. These two superconducting domes are unconventional and often share some common properties. The dome close to the ordered state has lower superconducting transition temperature ($T_{\rm{c}}$), and the higher $T_{\rm{c}}$ dome with non-Fermi liquid is far away from the ordered state \cite{twodome7,twodome8,twodome9}. Non-Fermi liquid and quantum critical point (QCP) are generally decoupled and well separated \cite{twodome10,twodome11}. These properties are summarized into a universal phase diagram and become the paradigm of the two superconducting domes in unconventional superconductors \cite{twodomeparadigm}.

Recently, it has been reported that the two superconducting domes are found in FeSe-based superconductors, which may have different pairing mechanisms \cite{FeSeTeQCP2,FeSeTeQCP}. Specifically, the superconducting dome SC1 near FeSe is mediated by the antiferromagnetic (AFM) fluctuations, whereas SC2 near FeSe$_{0.5}$Te$_{0.5}$ is associated with the pure nematic fluctuations. Although FeSe exhibits a nematic phase without long-range magnetic order below $\sim$90 K, several magnetic measurements have detected stripe-type AFM fluctuations within the nematic phase \cite{15AFM,FeSeSAFM}. With S doping and the application of pressure, the influence of the nematic fluctuations on Tc is limited, while spin fluctuations play a more significant role \cite{7,7-,12}. For SC2, elastoresistivity studies have demonstrated that FeSe1-xTex crosses a nonmagnetic pure nematic QCP, and the enhancement of superconductivity can be attributed to the nematic fluctuations \cite{FeSeTeQCP2,2016ubiquitous}. The shrinkage of the superconducting domes under high magnetic fields further supports that SC1 and SC2 are mediated by AFM fluctuations and nematic fluctuations, respectively \cite{FeSeTeQCP}. The paradigm and electronic structure of the two superconducting domes are crucial to the exploration of the unconventional superconducting pairing mechanism.

\begin{figure*}\centering
	\includegraphics[width=0.8\linewidth]{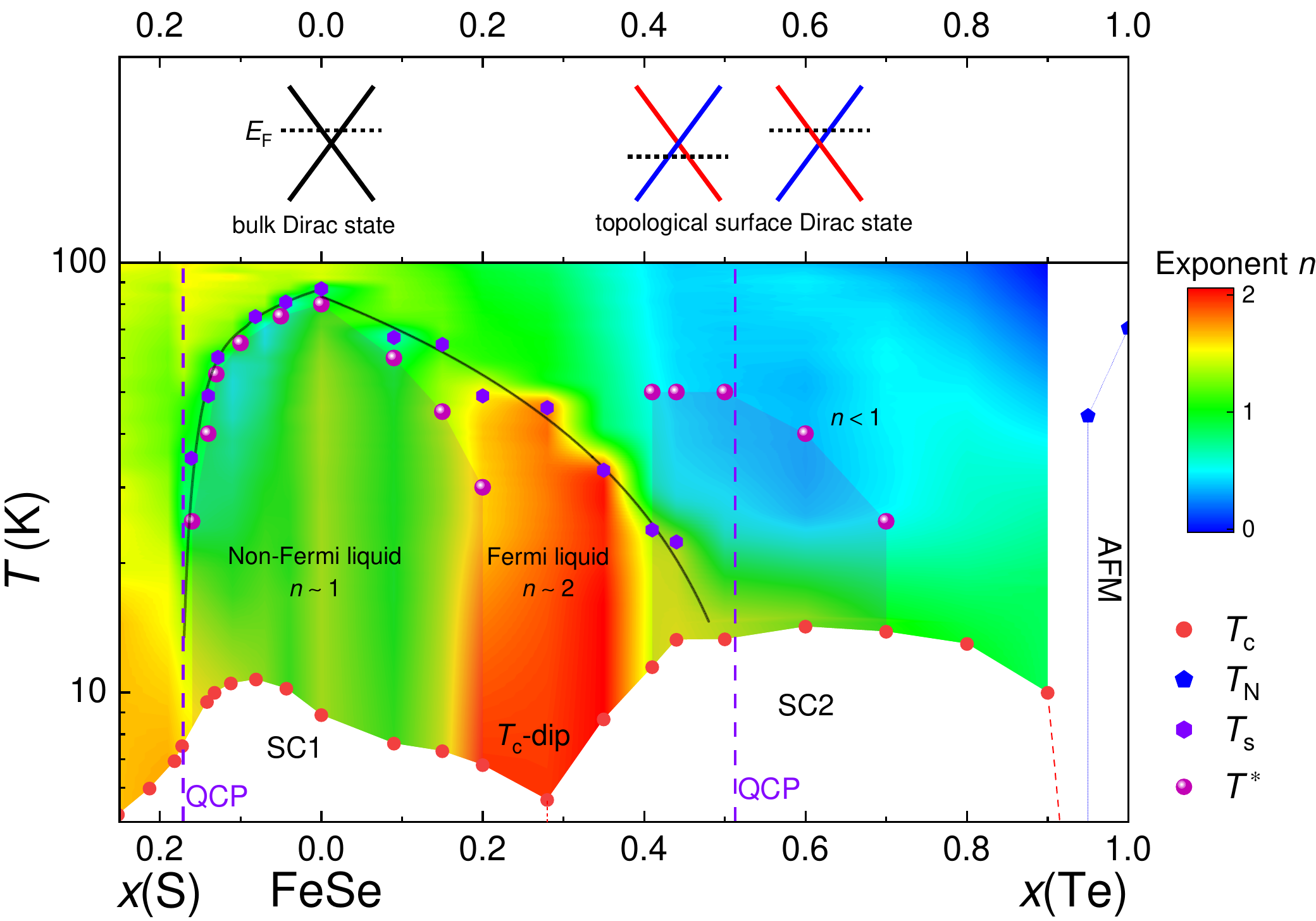}	
	\caption{Evolution phase diagram of the Dirac states, superconductivity, magnetic order and crystal structure with S and Te doping for FeSe$_{1-x}$S$_x$ and FeSe$_{1-x}$Te$_x$ single crystals. $T^{*}$ is the critical temperatures below which the Dirac states appear, $T\rm_{N}$ is the Néel temperature, $T\rm_{c}$ is the superconducting transition temperature, $T\rm_{s}$ is the structural (nematic) transition temperature. The temperature dependence of the exponent $n$ extracted from $d\rm{ln}(\rho - \rho_0)/ $$d\rm{ln} $$T$ for each crystal is represented as a contour plot. $T^{*}$, $T\rm_{c}$, $T\rm_{s}$ and $n$ in FeSe$_{1-x}$S$_x$ (0 $< x \le$ 0.25) are from ref.\cite{NFL7,NFL8,51sunFeSeS,FeSeShighfield,FeSeSAFM,FeSeSQCP}. The purple dashed line represents the nematic QCP near $x(\rm{S})\sim$ 0.17 \cite{FeSeSQCP} and $x(\rm{Te})\sim$ 0.52 \cite{FeSeTeQCP2,FeSeTeQCP}. Three schematic diagrams of Dirac states in top panel, namely the electron bulk Dirac state, the hole and electron topological surface Dirac states, in which the black dashed line is the Fermi surface, and the intersection of the red and blue solid lines represents the topological surface Dirac state characterized by the band inversion.}
\end{figure*}

FeSe is a compensated semimetal with the equal numbers of electron and hole carriers, in which hole pockets exist in the center of the Brillouin zone and electron pockets exist in the corners of the Brillouin zone \cite{FeSeARPES}. Especially in the nematic phase, the electron-type bulk Dirac cone located at the Brillouin zone corner was discovered \cite{FeSeDirac}. On the other hand, the normal state within the nematic phase of FeSe exhibits a good strange metal behavior, i.e. non-Fermi liquid \cite{NFL3, NFL8}. The study of FeSe$_{1-x}$S$_x$ under high fields suggests that the origin of the Dirac state may be related to strange metals \cite{NFL7}. Te substitution enhances the spin-orbit coupling (SOC), resulting in topologically non-trivial surface Dirac cones characterized by a band inversion along the $\Gamma$-Z line of the Brillouin zone, which has also been confirmed in magnetotransport measurements \cite{DiracFeSeTe,DiracFeSeTe2,39}. When the system becomes superconducting, these topological surface Dirac states could become gapped due to the proximity effect with the bulk superconductivity, leading to the formation of topological superconductivity [29]. Understanding the electronic structure is pivotal for unraveling the pairing mechanism. Due to challenges in sample preparation and measurement, previous research on the Dirac states has been confined to FeSe and the Te higher region. Consequently, the evolution from the bulk Dirac states to the topological surface Dirac states is still unclear. The study of the Dirac states and the normal state resistivity is crucial for the unconventional superconducting pairing mechanisms, particularly for SC2 associated with the pure nematic QCP.

\begin{figure*}\centering
	\includegraphics[width=1\linewidth]{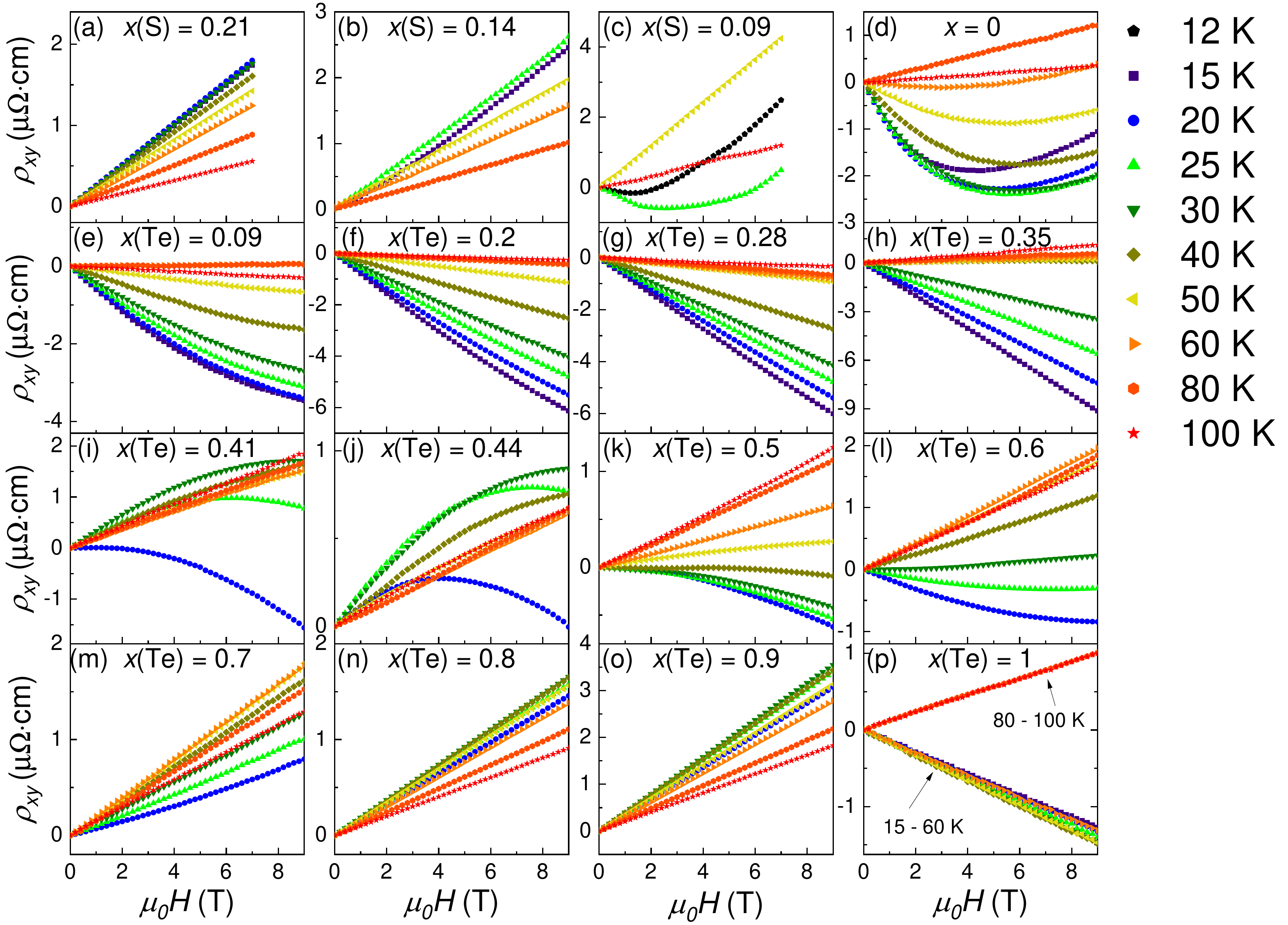}
	\caption{Magnetic field dependence of Hall resistivity $\rho_{xy}$ at different temperatures for FeSe$_{1-x}$S$_x$ (0 $<$ $x$ $\leq$ 0.21) and FeSe$_{1-x}$Te$_x$ (0 $\le x \le$ 1) single crystals. The data of FeSe$_{1-x}$S$_x$ (0 $<$ $x$ $\leq$ 0.21) are from refs. \cite{51sunFeSeS,FeSeSthreeband,nonfermi2020}.}\label{}
\end{figure*}

Electromagnetic transport stands as a convenient and efficient method for detecting superconductivity and electronic structure. In this study, high-quality FeSe$_{1-x}$S$_x$ (0 $\le x \le$ 0.25) FeSe$_{1-x}$Te$_x$ (0 $\le x \le$ 1) single crystals were successfully synthesized, with sizes reaching several mm$^2$. We have established a temperature-doping phase diagram, and observed the two superconducting domes, SC1 and SC2, as shown in Fig. 1 (See Figs. S1 and S2 for additional details). 
The normal state on SC1 far from the long-range AFM order at FeTe exhibits a non-Fermi liquid state, and the $T_{\rm{c}}$ is smaller than that of SC2 close to the AFM order. According to the comprehensive analysis of Hall resistivity and magnetoresistance (MR), we propose that there are three kinds of the Dirac states, the electron-type bulk Dirac state at the Brillouin zone corner, the hole- and electron-type surface Dirac states along the $\Gamma$-Z line. The topologically trivial bulk and topologically non-trivial surface Dirac states are accompanied by SC1 and SC2, respectively, suggesting that the origin of the two superconducting domes is due to the change in electronic structure, which is consistent with many previous researches \cite{twodomeFeS,twodomeCAC}. Our findings indicate that the two superconducting domes in FeSe-based superconductors are naturally the same as those in other unconventional superconductors, and the stark contrast between the Dirac states and the normal state resistivity of the two superconducting domes supports the possibility of two different pairing mechanisms.

FeSe$_{1-x}$Te$_x$ single crystals with 0 $\le$ $x$ $\le$ 0.5 were grown using the chemical vapor transport (CVT) method with a mixture of KCl/AlCl$_{3}$ as transport agents \cite{12}. FeSe$_{1-x}$Te$_x$ single crystals with 0.6 $\leq$ $x$ $\leq$ 1 were grown using self-flux method, and excess Fe was removed through Te-vapor annealing \cite{25Tevapor,sun_review_2019}. X-ray diffraction (XRD) measurements were conducted using a Rigaku X-ray diffractometer with Cu-K$\alpha$ radiation ($\lambda$ = 1.54 \AA). The elemental composition was determined using energy dispersive X-ray spectroscopy (EDX). Electrical and magnetic transport measurements were performed using physical property measurement system (PPMS-9T, Quantum Design). For magnetotranport measurements, a six-probe method was employed to simultaneously measure the Hall resistivity and MR. The magnetic field $H$ was applied parallel to the $c$ axis, perpendicular to the applied current. $H$ was scanned from -9 T to +9 T and the data was processed as follow, $\rho_{xy}$($H$) = [$\rho_{xy}$(+$H$) - $\rho_{xy}$(-$H$)]/2 and $\rho_{xx}$($H$) = [$\rho_{xx}$(+$H$) + $\rho_{xx}$(-$H$)]/2, which can effectively eliminate the longitudinal or transverse resistivity component caused by misalignment of the contacts.

\begin{figure*}\center
	\includegraphics[width=1\linewidth]{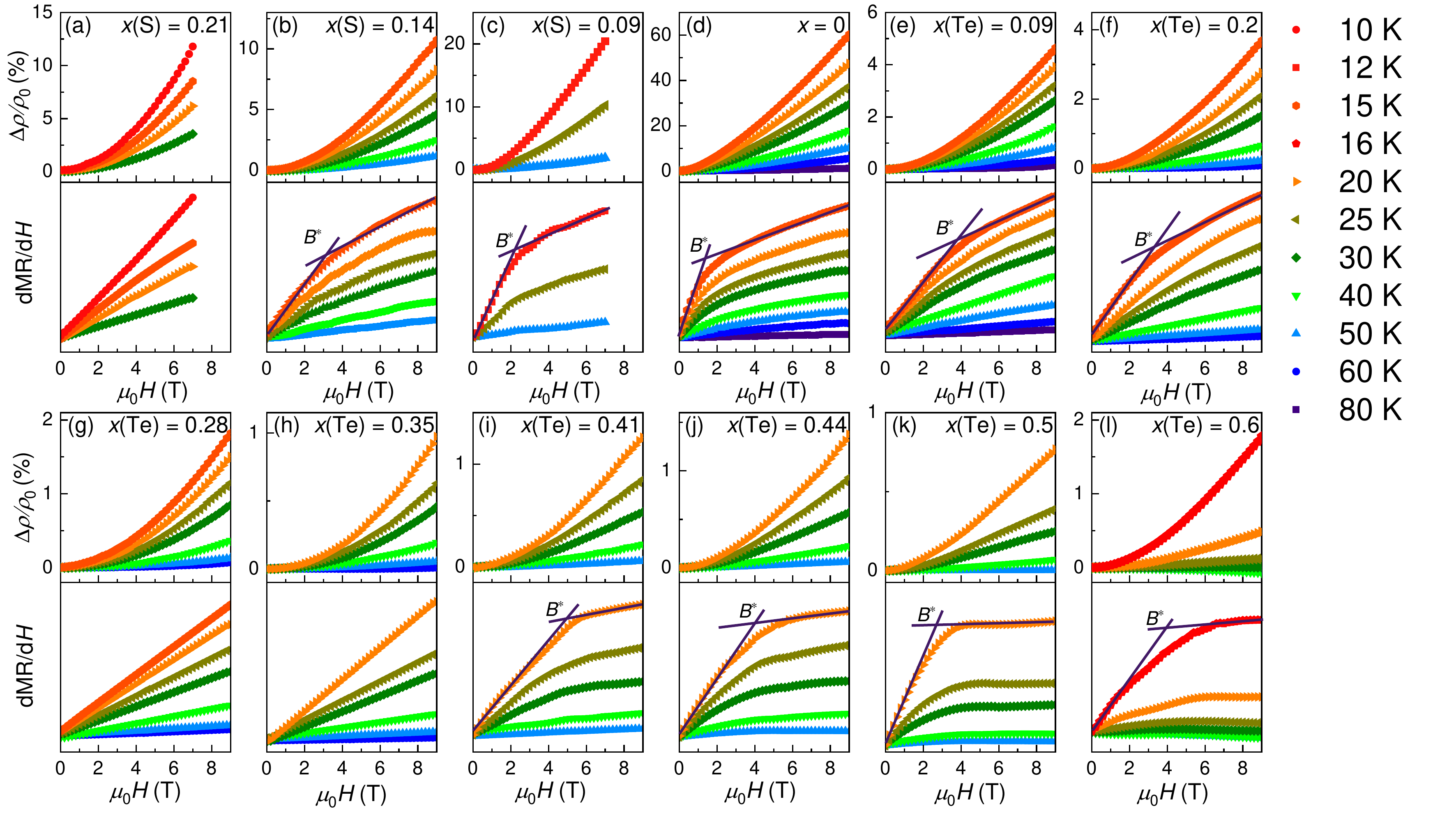}
	\caption{Magnetic field dependence of MR (Upper panel of (a-l)) and the first-order derivative of MR to $H$, d(MR)/d$H$ (Lower panel of (a-l)), at different temperatures for FeSe$_{1-x}$S$_x$ (0 $<$ $x$ $\leq$ 0.21) and FeSe$_{1-x}$Te$_x$ (0 $\leq$ $x$ $\leq$ 0.6) single crystals.	The data of FeSe$_{1-x}$S$_x$ (0 $<$ $x$ $\leq$ 0.21) are from refs. \cite{51sunFeSeS,FeSeSthreeband,nonfermi2020}.}\label{}
\end{figure*} 

Fig. 2 presents the magnetic field dependence of the Hall resistivity at various temperatures for FeSe$_{1-x}$S$_x$ (0 $\leq$ $x$ $\leq$ 0.21) and FeSe$_{1-x}$Te$_x$ (0 $\leq$ $x$ $\leq$ 1) single crystals. FeSe exhibits clear nonlinear behavior with a concave shape at low temperatures below 80 K, which has been ascribed to the emergence of the minority electron carriers with high mobility \cite{FeSemobilityspectra,FeSethreeband,FeSepressureHall}. The angle-resolved photoemission spectroscopy (ARPES) results of FeSe show that these minority electron bands with high mobility are the topologically trivial electronic bulk Dirac cones appearing around the Brillouin zone corner \cite{2,FeSeDirac,FeSeDiracmonolayer,FeSeDiracPRX2019}. This nonlinear behavior gradually diminishes with S or Te doping and becomes linear at $x$(S) = 0.21 and $x$(Te) = 0.28, indicating that the bulk Dirac states are gradually suppressed until they disappear. 

An interesting phenomenon occurs when the Te content increases to the range of 0.41$\sim$0.5: the nonlinear Hall behavior reappears, but with a convex shape. For $x$ = 0.6, the nonlinear Hall behavior becomes concave again, which is attributed to the electron Dirac state \cite{DiracTe0.6}. The ARPES results of FeSe$_{0.45}$Te$_{0.55}$ shows that, unlike FeSe, topologically nontrivial surface Dirac states are induced due to the enhancement of SOC by Te substitution \cite{DiracFeSeTe}. Compared to the concave shape in $x$ = 0.6, the convex nonlinear behavior can be naturally ascribed to the hole Dirac state based on the compensation effect in the whole FeSe$_{1-x}$Te$_x$ system. The compensated semimetal characteristic has been verified through various experiments, such as thermoelectric properties \cite{SeebeckFeSe,SeebeckFeTeSe}, ARPES \cite{39,37,38,34,36}, and Hall coefficients $R\rm_{H}$ at small fields (see Fig. S3). In order to provide the distribution of the Dirac states in FeSe-based superconductors, we define the characteristic temperature $T^{*}$ as the temperature at which nonlinear $\rho_{xy}$($H$) is observed just before the transition to linear behavior, representing the emergence of Dirac states (See Fig. S4 for the selection method). These $T^{*}$ values are summarized in Fig. 1. It should be emphasized that the variation in $T^{*}$ due to this selection has minimal impact on the phase diagram and the conclusions of our work. Moreover, at $x$ = 0.8 and 0.9, the Hall effects become linear, and Hall coefficients are positive without sign changes. In the case of FeTe, the Hall resistivity is linear, but the Hall coefficient changes from positive to negative below 70 K, which can be attributed to the antiferromagntic transition \cite{13}.

\begin{figure*}\center
	\includegraphics[width=1\linewidth]{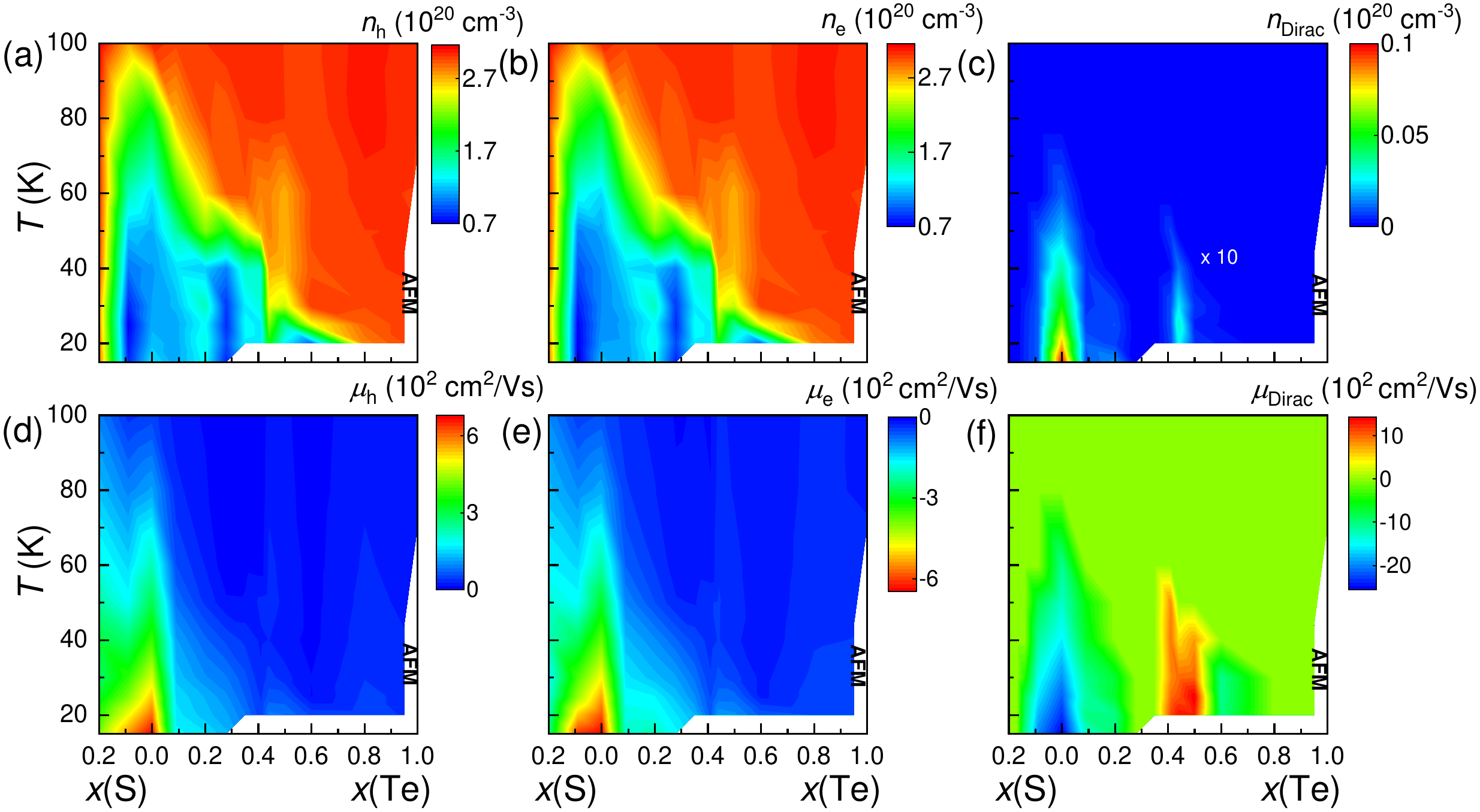}
	\caption{The cloud diagrams of the main carrier concentrations with (a) hole, (b) electron type, and (c) Dirac carrier concentrations, and their corresponding mobilities in (d-f) for FeSe$_{1-x}$S$_x$ (0 $< x \le$ 0.2) and FeSe$_{1-x}$Te$_x$ (0 $\le x \le$ 1) single crystals.The Dirac carrier concentration near FeSe$_{0.5}$Te$_{0.5}$ in (c) is magniffed 10 times. The mobility of hole carriers is deffned as positive, and the mobility of electron carriers is deffned as negative. The data of FeSe$_{1-x}$S$_x$ (0 $< x \le$ 0.2) are from refs.\cite{NFL7,FeSeSthreeband}}\label{}
\end{figure*}

It is well known that MR linearly dependent on magnetic field is one of the macroscopic manifestations of the Dirac state \cite{44,DiracMR}. In order to further verify the evolution of Dirac states with Te doping from Hall effects, we conducted MR measurements. Fig. 3 displays the magnetic field dependence of MR, defined as ($\rho(H)$-$\rho(0)$)/$\rho(0)$, and the first-order derivative of the MR to the magnetic field, d(MR)/d$H$, at different temperatures for FeSe$_{1-x}$S$_x$ (0 $\textless$ $x$ $\leq$ 0.21) and FeSe$_{1-x}$Te$_x$ (0 $\leq$ $x$ $\leq$ 0.6) single crystals. In the case of crystals with 0.7 $\leq$ $x$(Te) $\leq$ 1, the MR values are in the order of one thousandth and nearly zero, making the quantitative analysis unreliable. Therefore, we only analyze the MR data at 0 $\textless$ $x$(S) $\leq$  0.21 and 0 $\leq$ $x$(Te) $\leq$ 0.6. For FeSe, the MR value is approximately 60\% at 15 K and 9 T, and it gradually decreases as the temperature increases. However, the MR value is suppressed to less than 20\% for the S-doped crystals and less than 5\% for the Te-doped crystals due to the enhancement of scattering.

The first-order derivative of the MR to $H$ clearly demonstrates the field dependent behavior of MR. In the case of FeSe, the linear increase of d(MR)/d$H$ with magnetic field at low fields indicates a classic $B^{2}$ dependence of MR. The saturation of d(MR)/d$H$ to a much reduced slope above a characteristic field $B$* is typically attributed to the contribution of a linear field-dependent MR combined with a quadratic term. Linear MR is generally challenging to observe in normal materials because it requires large fields to reach the quantum limit, where all carriers occupy only the lowest Landau level (LL) \cite{40LL,41LL}. However, in certain materials hosting Dirac fermions with linear energy dispersion, linear MR can be easily detected in low or moderate fields, which has been verified in many iron-based materials, such as Ba(Sr)Fe$_2$As$_2$ \cite{44,45}, La(Pr)FeAsO \cite{46,47}, FeSe \cite{51sunFeSeS} and FeSe$_{0.6}$Te$_{0.4}$ \cite{DiracTe0.6}. 
	
In the case of FeSe$_{1-x}$S$_x$ and FeSe$_{1-x}$Te$_x$, the saturation of d(MR)/d$H$ is gradually suppressed with increasing $x$ and completely disappears at $x$(S) = 0.21 and $x$(Te) = 0.28. However, when the doping level reaches $x$(Te) = 0.41, the saturation of d(MR)/d$H$ reappears, and it maintains a constant value for 0.44 $\leq x$(Te) $\leq$ 0.6, indicating the dominance of Dirac states. Importantly, it should be noted that the region where linear MR is observed coincides with the region where nonlinear $\rho_{xy}$ is evident. The evolution of the Dirac states with Te doping from Hall resistivity has been further confirmed by MR. Moreover, the violation of the Kohler’s rule and the modified Kohler’s rule provide further evidences for the multiband effects of the FeSe$_{1-x}$Te$_x$ system, as shown in Figs. S5 and S6 \cite{SM}.

To investigate the multiband electronic structure of FeSe$_{1-x}$Te$_x$, and distinguish between the bulk Dirac and the topological surface Dirac states, quantitative analysis is necessary. A three-band model was utilized for the regions of 0 $\le x \le$ 0.2 and 0.41 $\le x \le$ 0.6 (low temperature regions with nonlinear Hall resistivity), considering two main carriers and one small carrier with high mobility. This model has been successfully applied to FeSe and confirmed by ARPES measurements \cite{FeSethreeband,FeSeARPES}. Additionally, a two-band model was employed to analyze the linear Hall resistivity and corresponding MR for the regions with linear Hall resistivity. The fitting process and typical results at 20 K (refer to Fig. S7 and Table S1 \cite{SM}) are shown in Supplementary Information \cite{SM}.

Fig. 4 illustrates the cloud diagram of carrier concentrations and mobility. The concentration of the main carriers, both hole and electron types, increases with temperature or S or Te doping, and remains nearly unchanged after the structural transition, as depicted in Figs. 4(a) and 4(b) for hole and electron carrier concentrations, respectively. Therefore, the structural transition can be attributed as the primary factor influencing the change of carrier concentration. On the other hand, the mobility of the main carriers is rapidly decreased in the low doping region, as shown in Figs. 4(d) and (e). For displaying the sign of the carriers, we define the mobility of hole carriers as positive and that of electron carriers as negative. Fig. 4(f) clearly shows the sign change behavior of the Dirac carriers with doping. The Dirac carrier concentration near $x$(Te) $\sim$ 0.5 is significantly smaller than that of FeSe (see Fig. 4(c), where the Dirac carrier concentration near $x$(Te) $\sim$ 0.5 is magnified 10 times), but the mobility is comparable to that of FeSe (see Fig. 4(f)). This observation is perfectly consistent with the fact of the bulk Dirac state around FeSe and the topological surface Dirac state around FeSe$_{0.5}$Te$_{0.5}$. The application of the three-band model has effectively quantitatively distinguished the bulk Dirac state from the topological surface Dirac state.

In order to illustrate the relationship between the transport properties, the Dirac states and two superconducting domes, a phase diagram was constructed, incorporating $T\rm_{c}$, the structural (nematic) transition temperature $T\rm_{s}$, the Néel temperature $T\rm_{N}$, $T^{*}$, and a contour plot of the power exponent $n$ extracted from $d\rm{ln}(\rho - \rho_0)/ $$d\rm{ln} $$T$ (see Fig. 1). In FeSe$_{1-x}$S$_x$ (0 $< x \le$ 0.25), the data points of $T^{*}$ were extracted from our previous work \cite{51sunFeSeS} and ref.\cite{NFL7}, and the other, including $T\rm_{c}$, $T\rm_{s}$ and $n$ are from ref.\cite{NFL8,FeSeShighfield,FeSeSAFM,FeSeSQCP,51sunFeSeS}. 
For FeSe, $T\rm_{s}$ and $T^{*}$ occur nearly simultaneously at approximately 90 K, suggesting that the bulk Dirac state at the Brillouin zone corner seems to originate from the nematic phase. However, analysis of the mobility spectrum in FeSe contradicts this speculation, as a remarkable reduction in carrier number and an enhancement in carrier mobility were simultaneously observed below 120 K but above $T\rm_{s}$, indicating the presence of another hidden order \cite{FeSemobilityspectra}. FeSe$_{1-x}$S$_x$ exhibits a non-Fermi liquid behavior within the nematic phase. $T\rm_{s}$ and $T^{*}$ also occur nearly simultaneously and disappear near the nematic QCP at $x (\rm{S})$ = 0.17. Magetotransport properties of FeSe$_{1-x}$S$_x$ suggests that the Dirac states can be ascribe to the strange metal component \cite{NFL7}. With Te doping, $T^{*}$ gradually deviates from $T\rm_{s}$ towards lower temperatures, disappearing at $x (\rm{Te})\sim$ 0.3, which corresponds to the junction of the two superconducting domes, forming a $T\rm_{c}$-dip. Meanwhile, in the nematic phase, the normal state gradually transitions from non-Fermi liquid to Fermi liquid. The bulk Dirac states completely overlap with the strange metal states, accompanied by the superconducting dome SC1. More importantly, strange metal behavior above the superconducting dome SC1 is similar to the normal state resistivity in iron pnictides and cuprates superconductors, supporting the superconducting pairing mechanism connected to the AFM fluctuations.

 The SC2 centers around the pure nematic QCP near $x\sim$ 0.52. When $x (\rm{Te})$ increases to $\sim$ 0.41, the topological surface Dirac state appears due to the enhancement of SOC. It is worth noting that $T^{*}$ is significantly higher than $T\rm_{s}$, which is opposite to that of lower S or Te doping. Furthermore, although $T\rm_{s}$ vanishes at $x (\rm{Te})\sim$ 0.5, $T^{*}$ persists until $x(\rm{Te})\sim$ 0.7. Fermi liquids also disappear with the reappearance of $T^{*}$, but the strange metals do not reappear, indicating that the topological surface Dirac state is different from the bulk Dirac state and has nothing to do with the strange metal. The rapid increase in $T\rm_{c}$ around $x$(Te) $\sim $ 0.4 with the emergence of the topological surface Dirac state may originate from the topological superconductivity induced by the proximity effect of the topological surface Dirac state and the ordinary $s$-wave superconductor \cite{ProximityEffect,18}. It is important that there is no non-Fermi liquid region above the SC2 similar to that above the SC1, which strongly supports the SC2 being associated with the QCP of pure electronic nematic order. The metallicity of the normal state gradually deteriorates with Te doping, i.e. $n <$ 1, which may be the normal state characteristic of the superconducting dome SC2 centered on the pure nematic QCP.
 
 Another interesting phenomenon is that the topological surface Dirac states change sign near the pure nematic QCP. There are three schematic diagrams of the Dirac states in the top panel of Fig. 1, showing the evolution of the Dirac states with S and Te doping. Near FeSe, the electron Dirac states at the Brillouin zone corner are bulk. With S or Te doping, the bulk Dirac state is gradually suppressed. Then, the topological surface Dirac state characterized by a band inversion appears around $x$(Te) $\sim $ 0.41 due to the enhancement of SOC. When crossing the nematic QCP around $x$(Te) $\sim $ 0.52 \cite{FeSeTeQCP,FeSeTeQCP2}, the surface Dirac state changes from hole to electron, and that is, the Fermi surface cross the Dirac point. The origin of the change is not yet clear, which may be related to the pure nematic QCP around $x$(Te) $\sim $ 0.52. The topological surface Dirac states and topological superconductivity are the keys for further studying the pure nematic QCP and superconducting pairing mechanism.
	
	Looking at the entire phase diagram, the normal state above SC1 far away from the long-range AFM order at FeTe exhibits a non-Fermi liquid state, but $T\rm_{c}$ is smaller than SC2 close to the AFM order, which is similar to the paradigm of the two superconducting domes in unconventional superconductors in Ref.\cite{twodomeparadigm}. However, the only difference is that in other unconventional superconductors, the superconducting dome with the non-Fermi liquid is generally higher than the other, and this anomaly in FeSe-based superconductors may be attributed to the topological superconductivity of SC2. The evolution of the Dirac states indicates that the appearance of the two superconducting domes may originate from the Fermi surface reconstruction with Te doping, similar to the quantum criticality transition in K$_{0.8}$Fe$_{2-x}$Se$_2$ \cite{twodomeQCT} and the Lifshitz transition in the heavily K-deposed FeSe films \cite{10}. The two superconducting domes have completely different normal state resistivity and Dirac states, strongly supporting two different superconducting pairing mechanisms.

In summary, we have successfully established the phase diagram with the two superconducting domes in FeSe-based superconductors, and provided the evolution of the Dirac states and normal state resistivity. Non-Fermi liquid appears on SC1, similar to iron pnictides and cuprates superconductors, supporting the superconducting pairing mechanism connected to AFM fluctuations. The power exponents above SC2 are less than 1, suggesting a potential signature of superconducting pairing mediated by pure nematic fluctuations. The performance of the two superconducting domes in FeSe-based superconductors is similar to the paradigm of the two superconducting domes found in many other unconventional superconductors, implying the same natural properties.

There are three kinds of Dirac states, the topologically trivial bulk Dirac state accompanied by the SC1 near FeSe, the hole and electron topologically non-trivial surface Dirac states accompanied by the SC2 near FeSe$_{0.5}$Te$_{0.5}$. The bulk Dirac states is closely related to the strange metal states and completely synchronized with them. The topological surface Dirac states undergo a sign change from hole to electron near the pure nematic QCP, providing important clues for understanding superconducting pairing mediated by pure nematic fluctuations. The junction of the two superconducting domes exhibits Fermi liquid behavior without Dirac state. The evolution of these Dirac states indicates that the appearance of the two superconducting domes may originate from the Fermi surface reconstruction. The two superconducting domes exhibit completely different Dirac and normal transport behaviors, strongly supporting the two distinct superconducting pairing mechanisms. 
	
\acknowledgements	
This work was partly supported by the National Key R$\& $D Program of China (Grant No. 2018YFA0704300), the National Natural Science Foundation of China (Grants No. 12374135, No. 12374136, No. 12204265, No. 12204487), the Fundamental Research Funds for the Central Universities (Grants No. 2242024k30029), the open research fund of Key Laboratory of Quantum Materials and Devices (Southeast University), Ministry of Education, and the Natural Science Foundation of Shandong Province (Grant No. ZR2022QA040). The authors thank Peng Zhang for useful discussions and the Center for Fundamental and Interdisciplinary Sciences of Southeast University for the support in XRD and physical property measurements.

Qiang Hou and Wei Wei contributed equally to this paper.

\bibliographystyle{unsrt}
\bibliography{ref}

\begin{thebibliography}{10}

\bibitem{twodomeCu}
G.~Grissonnanche, O.~Cyr-Choinière, F.~Laliberté, S.~René~de Cotret,
  A.~Juneau-Fecteau, S.~Dufour-Beauséjour, M.-È Delage, D.~LeBoeuf, J.~Chang,
  B.~J. Ramshaw, D.~A. Bonn, W.~N. Hardy, R.~Liang, S.~Adachi, N.~E. Hussey,
  B.~Vignolle, C.~Proust, M.~Sutherland, S.~Krämer, J.-H. Park, D.~Graf,
  N.~Doiron-Leyraud, and Louis Taillefer.
\newblock Direct measurement of the upper critical field in cuprate
  superconductors.
\newblock {\em Nat Commun}, 5:1--8, 2014.

\bibitem{twodomeCu214}
R.~A. Cooper, Y.~Wang, B.~Vignolle, O.~J. Lipscombe, S.~M. Hayden, Y.~Tanabe,
  T.~Adachi, Y.~Koike, M.~Nohara, H.~Takagi, Cyril Proust, and N.~E. Hussey.
\newblock Anomalous {Criticality} in the {Electrical} {Resistivity} of
  $\mathrm{La}_{2\ensuremath{-}x}\mathrm{Sr}_{x}\mathrm{CuO}_4$.
\newblock {\em Science}, 323(5914):603--607, 2009.

\bibitem{twodomeCu2141992}
Y.~Koike, A.~Kobayashi, T.~Kawaguchi, M.~Kato, T.~Noji, Y.~Ono, T.~Hikita, and
  Y.~Saito.
\newblock Anomalous $x$ dependence of ${T}_\mathrm{c}$ and possibility of
  low-temperature structural phase transition in
  $\mathrm{La}_{2\ensuremath{-}x}\mathrm{Sr}_{x}\mathrm{Cu}_{0.99}\mathrm{M}_{0.01}\mathrm{O}_4$
  ({M = Ni, Zn, Ga}).
\newblock {\em Solid State Communications}, 82(11):889--893, 1992.

\bibitem{twodomesun}
Liling Sun, Xiao-Jia Chen, Jing Guo, Peiwen Gao, Qing-Zhen Huang, Hangdong
  Wang, Minghu Fang, Xiaolong Chen, Genfu Chen, Qi~Wu, Chao Zhang, Dachun Gu,
  Xiaoli Dong, Lin Wang, Ke~Yang, Aiguo Li, Xi~Dai, Ho-kwang Mao, and Zhongxian
  Zhao.
\newblock Re-emerging superconductivity at 48 kelvin in iron chalcogenides.
\newblock {\em Nature}, 483(7387):67--69, 2012.

\bibitem{twodomeiron2014}
M.~Hiraishi, S.~Iimura, K.~M. Kojima, J.~Yamaura, H.~Hiraka, K.~Ikeda, P.~Miao,
  Y.~Ishikawa, S.~Torii, M.~Miyazaki, I.~Yamauchi, A.~Koda, K.~Ishii,
  M.~Yoshida, J.~Mizuki, R.~Kadono, R.~Kumai, T.~Kamiyama, T.~Otomo,
  Y.~Murakami, S.~Matsuishi, and H.~Hosono.
\newblock Bipartite magnetic parent phases in the iron oxypnictide
  superconductor.
\newblock {\em Nature Phys}, 10(4):300--303, 2014.

\bibitem{twodomeheavy}
H.~Q. Yuan, F.~M. Grosche, M.~Deppe, C.~Geibel, G.~Sparn, and F.~Steglich.
\newblock Observation of {Two} {Distinct} {Superconducting} {Phases} in
  $\mathrm{CeCu}_2\mathrm{Si}_2$.
\newblock {\em Science}, 302(5653):2104--2107, 2003.

\bibitem{twodome7}
Yoichi Ando, Seiki Komiya, Kouji Segawa, S.~Ono, and Y.~Kurita.
\newblock Electronic phase diagram of high-${T}_{c}$ cuprate superconductors
  from a mapping of the in-plane resistivity curvature.
\newblock {\em Phys. Rev. Lett.}, 93:267001, 2004.

\bibitem{twodome8}
Y.~Koike, T.~Kawaguchi, N.~Watanabe, T.~Noji, and Y.~Saito.
\newblock Superconductivity and low-temperature structural phase transition in
  {La$_{1.98-x}$Ce$_{0.02}$Ba$_x$CuO$_4$}.
\newblock {\em Solid State Communications}, 79(2):155--158, 1991.

\bibitem{twodome9}
Jie Yang, Rui Zhou, Lin-Lin Wei, Huai-Xin Yang, Jian-Qi Li, Zhong-Xian Zhao,
  and Guo-Qing Zheng.
\newblock New superconductivity dome in {LaFeAsO$_{1-x}$F$_x$} accompanied by
  structural transition.
\newblock {\em Chinese Physics Letters}, 32(10):107401, 2015.

\bibitem{twodome10}
B.~J. Ramshaw, S.~E. Sebastian, R.~D. McDonald, James Day, B.~S. Tan, Z.~Zhu,
  J.~B. Betts, Ruixing Liang, D.~A. Bonn, W.~N. Hardy, and N.~Harrison.
\newblock Quasiparticle mass enhancement approaching optimal doping in a
  high-${T}_\mathrm{c}$ superconductor.
\newblock {\em Science}, 348(6232):317--320, 2015.

\bibitem{twodome11}
H.~Luetkens, H.-H. Klauss, M.~Kraken, F.~J. Litterst, T.~Dellmann,
  R.~Klingeler, C.~Hess, R.~Khasanov, A.~Amato, C.~Baines, M.~Kosmala, O.~J.
  Schumann, M.~Braden, J.~Hamann-Borrero, N.~Leps, A.~Kondrat, G.~Behr,
  J.~Werner, and B.~Büchner.
\newblock The electronic phase diagram of the {LaO$_{1-x}$F$_x$FeAs}
  superconductor.
\newblock {\em Nature Mater}, 8(4):305--309, 2009.

\bibitem{twodomeparadigm}
Tanmoy Das and Christos Panagopoulos.
\newblock Two types of superconducting domes in unconventional superconductors.
\newblock {\em New Journal of Physics}, 18(10):103033, 2016.

\bibitem{FeSeTeQCP2}
Kousuke Ishida, Yugo Onishi, Masaya Tsujii, Kiyotaka Mukasa, Mingwei Qiu,
  Mikihiko Saito, Yuichi Sugimura, Kohei Matsuura, Yuta Mizukami, Kenichiro
  Hashimoto, and Takasada Shibauchi.
\newblock Pure nematic quantum critical point accompanied by a superconducting
  dome.
\newblock {\em Proceedings of the National Academy of Sciences},
  119(18):e2110501119, 2022.

\bibitem{FeSeTeQCP}
Kiyotaka Mukasa, Kousuke Ishida, Shusaku Imajo, Mingwei Qiu, Mikihiko Saito,
  Kohei Matsuura, Yuichi Sugimura, Supeng Liu, Yu~Uezono, Takumi Otsuka, Matija
  \ifmmode~\check{C}\else \v{C}\fi{}ulo, Shigeru Kasahara, Yuji Matsuda,
  Nigel~E. Hussey, Takao Watanabe, Koichi Kindo, and Takasada Shibauchi.
\newblock Enhanced superconducting pairing strength near a pure nematic quantum
  critical point.
\newblock {\em Phys. Rev. X}, 13:011032, 2023.

\bibitem{15AFM}
Qisi Wang, Yao Shen, Bingying Pan, Yiqing Hao, Mingwei Ma, Fang Zhou,
  P.~Steffens, K.~Schmalzl, T.~R. Forrest, M.~Abdel-Hafiez, Xiaojia Chen, D.~A.
  Chareev, A.~N. Vasiliev, P.~Bourges, Y.~Sidis, Huibo Cao, and Jun Zhao.
\newblock Strong interplay between stripe spin fluctuations, nematicity and
  superconductivity in {FeSe}.
\newblock {\em Nature Materials}, 15(2):159--163, 2016.

\bibitem{FeSeSAFM}
P.~Wiecki, K.~Rana, A.~E. B\"ohmer, Y.~Lee, S.~L. Bud'ko, P.~C. Canfield, and
  Y.~Furukawa.
\newblock Persistent correlation between superconductivity and
  antiferromagnetic fluctuations near a nematic quantum critical point in
  {FeSe$_{1-x}$S$_x$}.
\newblock {\em Phys. Rev. B}, 98:020507, 2018.

\bibitem{7}
J.~P. Sun, K.~Matsuura, G.~Z. Ye, Y.~Mizukami, M.~Shimozawa, K.~Matsubayashi,
  M.~Yamashita, T.~Watashige, S.~Kasahara, Y.~Matsuda, J.-Q. Yan, B.~C. Sales,
  Y.~Uwatoko, J.-G. Cheng, and T.~Shibauchi.
\newblock Dome-shaped magnetic order competing with high-temperature
  superconductivity at high pressures in {FeSe}.
\newblock {\em Nat Commun}, 7:12146, 2016.

\bibitem{7-}
K.~Matsuura, Y.~Mizukami, Y.~Arai, Y.~Sugimura, N.~Maejima, A.~Machida,
  T.~Watanuki, T.~Fukuda, T.~Yajima, Z.~Hiroi, K.~Y. Yip, Y.~C. Chan, Q.~Niu,
  S.~Hosoi, K.~Ishida, K.~Mukasa, S.~Kasahara, J.-G. Cheng, S.~K. Goh,
  Y.~Matsuda, Y.~Uwatoko, and T.~Shibauchi.
\newblock Maximizing \textit{{T}}$_{\textrm{c}}$ by tuning nematicity and
  magnetism in {FeSe}$_{\textrm{1-\textit{x}}}${S}$_{\textrm{\textit{x}}}$
  superconductors.
\newblock {\em Nature Communications}, 8(1):1143, 2017.

\bibitem{12}
K.~Mukasa, K.~Matsuura, M.~Qiu, M.~Saito, Y.~Sugimura, K.~Ishida, M.~Otani,
  Y.~Onishi, Y.~Mizukami, K.~Hashimoto, J.~Gouchi, R.~Kumai, Y.~Uwatoko, and
  T.~Shibauchi.
\newblock High-pressure phase diagrams of {FeSe$_{1-x}$Te$_x$}: correlation
  between suppressed nematicity and enhanced superconductivity.
\newblock {\em Nat Commun}, 12:1, 2021.

\bibitem{2016ubiquitous}
Hsueh-Hui Kuo, Jiun-Haw Chu, Johanna~C. Palmstrom, Steven~A. Kivelson, and
  Ian~R. Fisher.
\newblock Ubiquitous signatures of nematic quantum criticality in optimally
  doped {Fe}-based superconductors.
\newblock {\em Science}, 352(6288):958--962, 2016.

\bibitem{NFL7}
M.~\ifmmode~\check{C}\else \v{C}\fi{}ulo, M.~Berben, Y.-T. Hsu, J.~Ayres,
  R.~D.~H. Hinlopen, S.~Kasahara, Y.~Matsuda, T.~Shibauchi, and N.~E. Hussey.
\newblock Putative hall response of the strange metal component in
  $\mathrm{Fe}\mathrm{Se}_{1\ensuremath{-}x}\mathrm{S}_{x}$.
\newblock {\em Phys. Rev. Res.}, 3:023069, 2021.

\bibitem{NFL8}
Xiaolei Yi, Xiangzhuo Xing, Lingyao Qin, Jiajia Feng, Meng Li, Yufeng Zhang,
  Yan Meng, Nan Zhou, Yue Sun, and Zhixiang Shi.
\newblock Hydrothermal synthesis and complete phase diagram of
  $\mathrm{FeSe}_{1\ensuremath{-}x}\mathrm{S}_{x}
  (0\ensuremath{\le}x\ensuremath{\le}1)$ single crystals.
\newblock {\em Phys. Rev. B}, 103:144501, 2021.

\bibitem{51sunFeSeS}
Yue Sun, Sunseng Pyon, and Tsuyoshi Tamegai.
\newblock Electron carriers with possible {Dirac}-cone-like dispersion in
  {${\mathrm{FeSe}}_{1\ensuremath{-}x}{\mathrm{S}}_{x}$}
  ($x\phantom{\rule{4pt}{0ex}}=0$ and 0.14) single crystals triggered by
  structural transition.
\newblock {\em Phys. Rev. B}, 93:104502, 2016.

\bibitem{FeSeShighfield}
M.~Bristow, P.~Reiss, A.~A. Haghighirad, Z.~Zajicek, S.~J. Singh, T.~Wolf,
  D.~Graf, W.~Knafo, A.~McCollam, and A.~I. Coldea.
\newblock Anomalous high-magnetic field electronic state of the nematic
  superconductors {FeSe$_{1-x}$S$_x$}.
\newblock {\em Phys. Rev. Res.}, 2:013309, 2020.

\bibitem{FeSeSQCP}
Suguru Hosoi, Kohei Matsuura, Kousuke Ishida, Hao Wang, Yuta Mizukami, Tatsuya
  Watashige, Shigeru Kasahara, Yuji Matsuda, and Takasada Shibauchi.
\newblock Nematic quantum critical point without magnetism in
  {FeSe$_{1-x}$S$_x$} superconductors.
\newblock {\em Proceedings of the National Academy of Sciences},
  113(29):8139--8143, 2016.

\bibitem{FeSeARPES}
M.~D. Watson, T.~K. Kim, A.~A. Haghighirad, N.~R. Davies, A.~McCollam,
  A.~Narayanan, S.~F. Blake, Y.~L. Chen, S.~Ghannadzadeh, A.~J. Schofield,
  M.~Hoesch, C.~Meingast, T.~Wolf, and A.~I. Coldea.
\newblock Emergence of the nematic electronic state in {FeSe}.
\newblock {\em Phys. Rev. B}, 91:155106, 2015.

\bibitem{FeSeDirac}
S.~Y. Tan, Y.~Fang, D.~H. Xie, W.~Feng, C.~H.~P. Wen, Q.~Song, Q.~Y. Chen,
  W.~Zhang, Y.~Zhang, L.~Z. Luo, B.~P. Xie, X.~C. Lai, and D.~L. Feng.
\newblock Observation of {Dirac} cone band dispersions in {FeSe} thin films by
  photoemission spectroscopy.
\newblock {\em Phys. Rev. B}, 93:104513, 2016.

\bibitem{NFL3}
S.~Licciardello, J.~Buhot, J.~Lu, J.~Ayres, S.~Kasahara, Y.~Matsuda,
  T.~Shibauchi, and N.~E. Hussey.
\newblock Electrical resistivity across a nematic quantum critical point.
\newblock {\em Nature}, 567(7747), 2019.

\bibitem{DiracFeSeTe}
Peng Zhang, Koichiro Yaji, Takahiro Hashimoto, Yuichi Ota, Takeshi Kondo, Kozo
  Okazaki, Zhijun Wang, Jinsheng Wen, G.~D. Gu, Hong Ding, and Shik Shin.
\newblock Observation of topological superconductivity on the surface of an
  iron-based superconductor.
\newblock {\em Science}, 360(6385):182--186, 2018.

\bibitem{DiracFeSeTe2}
Peng Zhang, Zhijun Wang, Xianxin Wu, Koichiro Yaji, Yukiaki Ishida, Yoshimitsu
  Kohama, Guangyang Dai, Yue Sun, Cedric Bareille, Kenta Kuroda, Takeshi Kondo,
  Kozo Okazaki, Koichi Kindo, Xiancheng Wang, Changqing Jin, Jiangping Hu,
  Ronny Thomale, Kazuki Sumida, Shilong Wu, Koji Miyamoto, Taichi Okuda, Hong
  Ding, G.~D. Gu, Tsuyoshi Tamegai, Takuto Kawakami, Masatoshi Sato, and Shik
  Shin.
\newblock Multiple topological states in iron-based superconductors.
\newblock {\em Nature Phys}, 15(1):41--47, 2019.

\bibitem{39}
Zhijun Wang, P.~Zhang, Gang Xu, L.~K. Zeng, H.~Miao, Xiaoyan Xu, T.~Qian,
  Hongming Weng, P.~Richard, A.~V. Fedorov, H.~Ding, Xi~Dai, and Zhong Fang.
\newblock Topological nature of the $\mathrm{FeSe}_{0.5}\mathrm{Te}_{0.5}$
  superconductor.
\newblock {\em Phys. Rev. B}, 92:115119, 2015.

\bibitem{FeSeSthreeband}
Y~A Ovchenkov, D~A Chareev, V~A Kulbachinskii, V~G Kytin, D~E Presnov, O~S
  Volkova, and A~N Vasiliev.
\newblock Highly mobile carriers in iron-based superconductors.
\newblock {\em Superconductor Science and Technology}, 30(3):035017, 2017.

\bibitem{nonfermi2020}
W.~K. Huang, S.~Hosoi, M.~Čulo, S.~Kasahara, Y.~Sato, K.~Matsuura,
  Y.~Mizukami, M.~Berben, N.~E. Hussey, H.~Kontani, T.~Shibauchi, and
  Y.~Matsuda.
\newblock Non-{Fermi} liquid transport in the vicinity of the nematic quantum
  critical point of superconducting {FeSe$_{1-x}$S$_x$}.
\newblock {\em Physical Review Research}, 2(3):033367, 2020.

\bibitem{twodomeFeS}
Makoto Shimizu, Nayuta Takemori, Daniel Guterding, and Harald~O. Jeschke.
\newblock Two-dome superconductivity in {FeS} induced by a {Lifshitz}
  transition.
\newblock {\em Phys. Rev. Lett.}, 121:137001, 2018.

\bibitem{twodomeCAC}
J.~P. Sun, P.~Shahi, H.~X. Zhou, Y.~L. Huang, K.~Y. Chen, B.~S. Wang, S.~L. Ni,
  N.~N. Li, K.~Zhang, W.~G. Yang, Y.~Uwatoko, G.~Xing, J.~Sun, D.~J. Singh,
  K.~Jin, F.~Zhou, G.~M. Zhang, X.~L. Dong, Z.~X. Zhao, and J.-G. Cheng.
\newblock Reemergence of high-${T}_\mathrm{c}$ superconductivity in the
  ({Li}$_{1-x}${Fe}$_x$){OHFe}$_{1-y}${Se} under high pressure.
\newblock {\em Nat Commun}, 9(1):380, 2018.

\bibitem{25Tevapor}
Yue Sun, Yuji Tsuchiya, Tatsuhiro Yamada, Toshihiro Taen, Sunseng Pyon,
  Zhixiang Shi, and Tsuyoshi Tamegai.
\newblock Evolution of superconductivity in {Fe$_{1+y}$Te$_{1-x}$Se$_x$}
  annealed in {Te} vapor.
\newblock {\em J. Phys. Soc. Jpn.}, 82:093705, 2013.

\bibitem{sun_review_2019}
Yue Sun, Zhixiang Shi, and Tsuyoshi Tamegai.
\newblock Review of annealing effects and superconductivity in
  {FeSe$_{1-x}$Te$_x$} superconductors.
\newblock {\em Supercond. Sci. Technol.}, 32:103001, 2019.

\bibitem{FeSemobilityspectra}
K.~K. Huynh, Y.~Tanabe, T.~Urata, H.~Oguro, S.~Heguri, K.~Watanabe, and
  K.~Tanigaki.
\newblock Electric transport of a single-crystal iron chalcogenide {FeSe}
  superconductor: Evidence of symmetry-breakdown nematicity and additional
  ultrafast {Dirac} cone-like carriers.
\newblock {\em Phys. Rev. B}, 90:144516, 2014.

\bibitem{FeSethreeband}
M.~D. Watson, T.~Yamashita, S.~Kasahara, W.~Knafo, M.~Nardone, J.~B\'eard,
  F.~Hardy, A.~McCollam, A.~Narayanan, S.~F. Blake, T.~Wolf, A.~A. Haghighirad,
  C.~Meingast, A.~J. Schofield, H.~v.~L\"ohneysen, Y.~Matsuda, A.~I. Coldea,
  and T.~Shibauchi.
\newblock Dichotomy between the hole and electron behavior in multiband
  superconductor {FeSe} probed by ultrahigh magnetic fields.
\newblock {\em Phys. Rev. Lett.}, 115:027006, 2015.

\bibitem{FeSepressureHall}
J.~P. Sun, G.~Z. Ye, P.~Shahi, J.-Q. Yan, K.~Matsuura, H.~Kontani, G.~M. Zhang,
  Q.~Zhou, B.~C. Sales, T.~Shibauchi, Y.~Uwatoko, D.~J. Singh, and J.-G. Cheng.
\newblock High-${T}_{c}$ superconductivity in {FeSe} at high pressure: Dominant
  hole carriers and enhanced spin fluctuations.
\newblock {\em Phys. Rev. Lett.}, 118:147004, 2017.

\bibitem{2}
Takasada Shibauchi, Tetsuo Hanaguri, and Yuji Matsuda.
\newblock Exotic superconducting states in {FeSe}-based materials.
\newblock {\em J. Phys. Soc. Jpn.}, 89:102002, 2020.

\bibitem{FeSeDiracmonolayer}
S.~Kanayama, K.~Nakayama, G.~N. Phan, M.~Kuno, K.~Sugawara, T.~Takahashi, and
  T.~Sato.
\newblock Two-dimensional {Dirac} semimetal phase in undoped one-monolayer
  {FeSe} film.
\newblock {\em Phys. Rev. B}, 96:220509, 2017.

\bibitem{FeSeDiracPRX2019}
M.~Yi, H.~Pfau, Y.~Zhang, Y.~He, H.~Wu, T.~Chen, Z.~R. Ye, M.~Hashimoto, R.~Yu,
  Q.~Si, D.-H. Lee, Pengcheng Dai, Z.-X. Shen, D.~H. Lu, and R.~J. Birgeneau.
\newblock Nematic energy scale and the missing electron pocket in {FeSe}.
\newblock {\em Phys. Rev. X}, 9:041049, 2019.

\bibitem{DiracTe0.6}
Yue Sun, Toshihiro Taen, Tatsuhiro Yamada, Sunseng Pyon, Terukazu Nishizaki,
  Zhixiang Shi, and Tsuyoshi Tamegai.
\newblock Multiband effects and possible dirac fermions in
  {${\text{Fe}}_{1+y}$${\text{Te}}_{0.6}$${\text{Se}}_{0.4}$}.
\newblock {\em Phys. Rev. B}, 89:144512, 2014.

\bibitem{SeebeckFeSe}
T.~M. McQueen, Q.~Huang, V.~Ksenofontov, C.~Felser, Q.~Xu, H.~Zandbergen, Y.~S.
  Hor, J.~Allred, A.~J. Williams, D.~Qu, J.~Checkelsky, N.~P. Ong, and R.~J.
  Cava.
\newblock Extreme sensitivity of superconductivity to stoichiometry in
  {${\text{Fe}}_{1+\ensuremath{\delta}}\text{Se}$}.
\newblock {\em Phys. Rev. B}, 79:014522, 2009.

\bibitem{SeebeckFeTeSe}
I.~Pallecchi, G.~Lamura, M.~Tropeano, M.~Putti, R.~Viennois, E.~Giannini, and
  D.~Van~der Marel.
\newblock Seebeck effect in
  {${\text{Fe}}_{1+x}{\text{Te}}_{1\ensuremath{-}y}{\text{Se}}_{y}$} single
  crystals.
\newblock {\em Phys. Rev. B}, 80:214511, 2009.

\bibitem{37}
Z.~K. Liu, M.~Yi, Y.~Zhang, J.~Hu, R.~Yu, J.-X. Zhu, R.-H. He, Y.~L. Chen,
  M.~Hashimoto, R.~G. Moore, S.-K. Mo, Z.~Hussain, Q.~Si, Z.~Q. Mao, D.~H. Lu,
  and Z.-X. Shen.
\newblock Experimental observation of incoherent-coherent crossover and
  orbital-dependent band renormalization in iron chalcogenide superconductors.
\newblock {\em Phys. Rev. B}, 92:235138, 2015.

\bibitem{38}
Jianwei Huang, Rong Yu, Zhijun Xu, Jian-Xin Zhu, Ji~Seop Oh, Qianni Jiang, Meng
  Wang, Han Wu, Tong Chen, Jonathan~D. Denlinger, Sung-Kwan Mo, Makoto
  Hashimoto, Matteo Michiardi, Tor~M. Pedersen, Sergey Gorovikov, Sergey
  Zhdanovich, Andrea Damascelli, Genda Gu, Pengcheng Dai, Jiun-Haw Chu, Donghui
  Lu, Qimiao Si, Robert~J. Birgeneau, and Ming Yi.
\newblock Correlation-driven electronic reconstruction in {FeTe$_{1-x}$Se$_x$}.
\newblock {\em Commun Phys}, 5:1, 2022.

\bibitem{34}
K.~Okazaki, Y.~Ito, Y.~Ota, Y.~Kotani, T.~Shimojima, T.~Kiss, S.~Watanabe,
  C.~T. Chen, S.~Niitaka, T.~Hanaguri, H.~Takagi, A.~Chainani, and S.~Shin.
\newblock Evidence for a $\mathrm{cos}﻿(4\ensuremath{\varphi})$ modulation of
  the superconducting energy gap of optimally doped
  {${\mathrm{FeTe}}_{0.6}{\mathrm{Se}}_{0.4}$} single crystals using laser
  angle-resolved photoemission spectroscopy.
\newblock {\em Phys. Rev. Lett.}, 109:237011, 2012.

\bibitem{36}
Fei Chen, Bo~Zhou, Yan Zhang, Jia Wei, Hong-Wei Ou, Jia-Feng Zhao, Cheng He,
  Qing-Qin Ge, Masashi Arita, Kenya Shimada, Hirofumi Namatame, Masaki
  Taniguchi, Zhong-Yi Lu, Jiangping Hu, Xiao-Yu Cui, and D.~L. Feng.
\newblock Electronic structure of
  {${\text{Fe}}_{1.04}{\text{Te}}_{0.66}{\text{Se}}_{0.34}$}.
\newblock {\em Phys. Rev. B}, 81:014526, 2010.

\bibitem{13}
Yue Sun, Tatsuhiro Yamada, Sunseng Pyon, and Tsuyoshi Tamegai.
\newblock Influence of interstitial {Fe} to the phase diagram of
  {Fe$_{1+y}$Te$_{1-x}$Se$_x$} single crystals.
\newblock {\em Sci Rep}, 6:1, 2016.

\bibitem{44}
Khuong~K. Huynh, Yoichi Tanabe, and Katsumi Tanigaki.
\newblock Both electron and hole {Dirac} cone states in
  $\mathrm{Ba}(\mathrm{FeAs}{)}_{2}$ confirmed by magnetoresistance.
\newblock {\em Phys. Rev. Lett.}, 106:217004, 2011.

\bibitem{DiracMR}
Tian Liang, Quinn Gibson, Mazhar~N. Ali, Minhao Liu, R.~J. Cava, and N.~P. Ong.
\newblock Ultrahigh mobility and giant magnetoresistance in the {Dirac}
  semimetal {Cd$_{3}$As$_{2}$}.
\newblock {\em Nature Mater}, 14(3):280--284, 2015.

\bibitem{40LL}
A.~A. Abrikosov.
\newblock Quantum magnetoresistance.
\newblock {\em Phys. Rev. B}, 58:2788, 1998.

\bibitem{41LL}
A.~A. Abrikosov.
\newblock Quantum linear magnetoresistance.
\newblock {\em Europhysics Letters}, 49:789, 2000.

\bibitem{45}
S.~V. Chong, G.~V.~M. Williams, J.~Kennedy, F.~Fang, J.~L. Tallon, and
  K.~Kadowaki.
\newblock Large low-temperature magnetoresistance in
  $\mathrm{Sr}\mathrm{Fe_{2}As_{2}}$ single crystals.
\newblock {\em Europhysics Letters}, 104:17002, 2013.

\bibitem{46}
I.~Pallecchi, F.~Bernardini, M.~Tropeano, A.~Palenzona, A.~Martinelli,
  C.~Ferdeghini, M.~Vignolo, S.~Massidda, and M.~Putti.
\newblock Magnetotransport in {La(Fe,Ru)AsO} as a probe of band structure and
  mobility.
\newblock {\em Phys. Rev. B}, 84:134524, 2011.

\bibitem{47}
D.~Bhoi, P.~Mandal, P.~Choudhury, S.~Pandya, and V.~Ganesan.
\newblock {Quantum magnetoresistance of the PrFeAsO oxypnictide}.
\newblock {\em Appl. Phys. Lett.}, 98:172105, 2011.

\bibitem{SM}
See supplementary information.

\bibitem{ProximityEffect}
Liang Fu and C.~L. Kane.
\newblock Superconducting proximity effect and majorana fermions at the surface
  of a topological insulator.
\newblock {\em Phys. Rev. Lett.}, 100:096407, 2008.

\bibitem{18}
Dongfei Wang, Lingyuan Kong, Peng Fan, Hui Chen, Shiyu Zhu, Wenyao Liu, Lu~Cao,
  Yujie Sun, Shixuan Du, John Schneeloch, Ruidan Zhong, Genda Gu, Liang Fu,
  Hong Ding, and Hong-Jun Gao.
\newblock Evidence for majorana bound states in an iron-based superconductor.
\newblock {\em Science}, 362:333, 2018.

\bibitem{twodomeQCT}
Jing Guo, Xiao-Jia Chen, Jianhui Dai, Chao Zhang, Jiangang Guo, Xiaolong Chen,
  Qi~Wu, Dachun Gu, Peiwen Gao, Lihong Yang, Ke~Yang, Xi~Dai, Ho-kwang Mao,
  Liling Sun, and Zhongxian Zhao.
\newblock Pressure-driven quantum criticality in iron-selenide superconductors.
\newblock {\em Phys. Rev. Lett.}, 108:197001, 2012.

\bibitem{10}
X.~Shi, Z.-Q. Han, X.-L. Peng, P.~Richard, T.~Qian, X.-X. Wu, M.-W. Qiu, S.~C.
  Wang, J.~P. Hu, Y.-J. Sun, and H.~Ding.
\newblock Enhanced superconductivity accompanying a {Lifshitz} transition in
  electron-doped {FeSe} monolayer.
\newblock {\em Nat Commun}, 8:14988, 2017.

\end{thebibliography}

	\pagebreak
	
	\newpage
	\onecolumngrid
	\begin{center}
		\textbf{\huge Supplemental information}
	\end{center}
	\vspace{1cm}
	\twocolumngrid
	\setcounter{equation}{0}
	\setcounter{figure}{0}
	\setcounter{table}{0}
	
	\makeatletter
	\renewcommand{\theequation}{S\arabic{equation}}
	\renewcommand{\thefigure}{S\arabic{figure}}
	
	\section{Kohler's rule}
Kohler’s rule is one of the powerful means to test whether the electronic structure of a sample belongs to single band. According to the Boltzmann transport theory, the MR at different temperatures, in the Fermi liquid state of a single-band system with isotropic scattering, can be scaled by the Kohler’s rule, which can be described as the common form $\Delta \rho_{xx}(H)/\rho_{xx}(0) = f[H/\rho_{xx}(0)]$ [1,2]. The MR at different temperatures for FeSe$_{1-x}$Te$_x$ (0 $\le x \le$ 0.6) plotted as a function of $(\mu_0 H/\rho_{xx}(0))^2$ is shown in Fig. S2. Obviously, the MR data cannot be scaled, indicating that no sample follows the Kohler’s rule. However, although the Kohler’s rule was violated in some cuprate and iron-based superconductors due to antiferromagnetism fluctuations, MR can be scaled by the modified Kohler’ rule, $\Delta \rho_{xx}(H)/\rho_{xx}(0) \propto  {\rm{tan}}^2 \theta_H = [\rho_{xy}(H)/\rho_{xx}(0)]^2$ [3–5]. Similarly, MR at different temperature for FeSe$_{1-x}$Te$_x$ (0 $\le x \le$ 0.6) plotted as a function of $\rm{tan}^2 \theta_H$ is shown in Figure S3 and the modified Kohler’s rule is also violated. So, the violations of Kohler and modified Kohler’s rule strongly prove the multi-bands effects in this system.

\section{Three band model}

The magnetic field dependence of conductivity tensor components can be expressed as:

\begin{equation}
	\sigma_{xx}=\sum_{i=1}^{l} \frac{\sigma_{i}}{1+\mu_i^2 B^2},\label{1}
\end{equation}
\begin{equation}
	\sigma_{xy}=\sum_{i=1}^{l} \frac{s_i \sigma_{i}\mu_i B}{1+\mu_i^2 B^2},\label{2}
\end{equation}
\begin{equation}
	\sigma_{i}=en_i\mu_i,\label{1}
\end{equation}
where $\sigma_{i}$ are conductivity tensor components, $i$ is a band index, $\mu_i$ is a carrier mobility, $n_i$ is a carrier concentration, $s_i$ is "-1" for a hole and "+1" for an electron bands. Resistivity tensor in tetragonal phase can be derived as follow:
\begin{equation}
	\rho_{xx}=\rho_{yy}=\frac{\sigma_{xx}}{\sigma_{xx}^2+\sigma_{xy}^2},\label{2}
\end{equation}
\begin{equation}
	-\rho_{xy}=\rho_{yx}=\frac{\sigma_{xy}}{\sigma_{xx}^2+\sigma_{xy}^2}.\label{2}
\end{equation}

Based on the analysis of Hall and MR, certain constraints were applied during the three-band fitting. These constraints are as follows: 

\setlength{\hangindent}{1.4em}\noindent 1.	Due to the compensation of the system, the carrier concentrations of electrons and holes remain equal at all times.

\setlength{\hangindent}{1.4em}\noindent 2.	The carrier concentration of FeSe is in the order of 10$^{20}$ $\rm{cm}^{-3}$. Since Te substitution is equivalent doping, this initial value is used for fitting of the whole system.

\setlength{\hangindent}{1.4em}\noindent 3.	The concave shape of hall behavior represents two electron bands and one hole band, while the convex shape of hall behavior represents one electron band and two hole bands. 

Thus, we have carried out three band fitting for Hall behavior and corresponding MR at the same time. Fig. S7 shows the fitting results at 20 K, which well explains the nonlinear Hall behavior and reveals the existence of small bands with high mobility. Therefore, our analysis shows that the Fermi surface of the system is complex and very sensitive to Te content. 

Table 1 gives the information about carrier concentration and mobility at 20 K, obtained by three band fitting. With the change of Te content, there is no significant change in the concentration of main carriers, which proves an equivalent doping effect of Te replacing Se; However, the mobility decreases significantly, possibly due to the enhancement of lattice scattering caused by the larger radius of Te ions. Furthermore, it is intriguing to explore the changes in the type of small band with Te doping. It is possible that the introduction of Te causes a slight reconstruction of the Fermi surface, resulting in the generation of different types of small bands. Additionally, the presence of small bands with high mobility for hole-type carriers may be closely related to nematic fluctuations. The underlying reasons for these observations warrant further investigation and study.
	
	\section{Two band model}
	In order to reveal the evolution of carrier concentration and mobility with temperature and Te content in the whole region, we also use the compensated two band model to fit the linear Hall behavior [6].
	\begin{equation}
		\rho_{xx}(0)=\frac{1}{ne}\frac{1}{\mu_h+\mu_e},\label{2}
	\end{equation}
	\begin{equation}
		\rho_{xy}(H)=\frac{1}{ne}\frac{\mu_h-\mu_e}{\mu_h+\mu_e}\mu_0H,\label{2}
	\end{equation}
	\begin{equation}
		\frac{\Delta\rho_{xx}(H)}{\rho_{xx}(0)}=\mu_h\mu_e(\mu H)^2.\label{2}
	\end{equation}
	
	where $\mu_h(\mu_e)$ is the hole (electron) carrier mobility, $n_h(n_e)$ is the hole carrier concentration. In the high-Te region where the MR is nearly zero, we have observed that the Hall coefficients remain relatively unchanged with varying Te content. Based on this observation, we can reasonably assume that the carrier concentration does not undergo significant changes and is likely to be close to the concentration found in the adjacent region.
	
	\begin{figure*}\
		\includegraphics[width=1\linewidth]{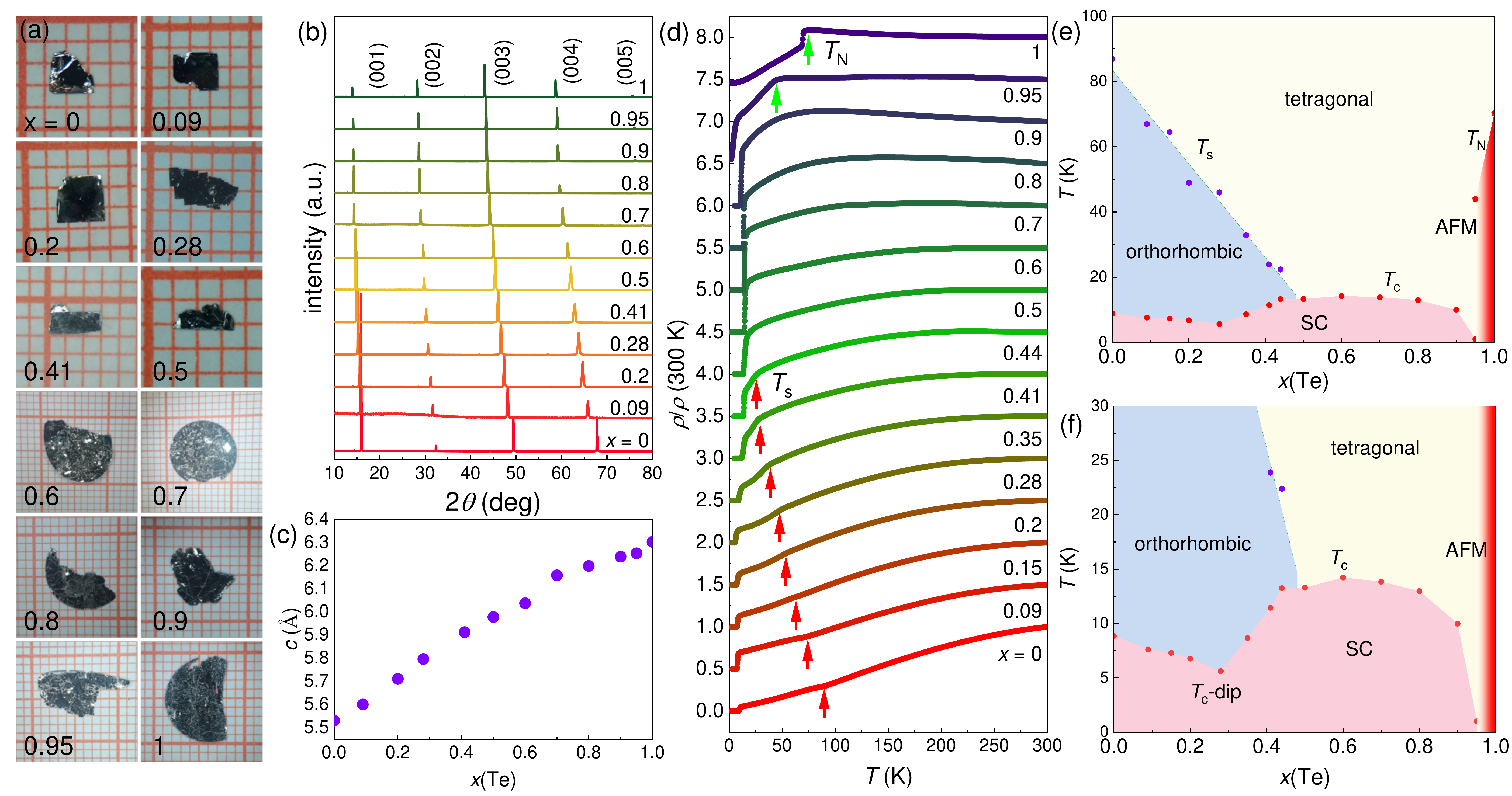}
		\caption{(a) Optical image of FeSe$_{1-x}$Te$_x$ (0 $\leq x\leq$ 1) single crystals, and the actual Se/Te content determined using EDX; (b) X-ray diffractions patterns of FeSe$_{1-x}$Te$_x$ (0 $\leq x\leq$ 1) single crystals along the $c$ axis; (c) Monotonic increase of the $c$-axis length on substitution $x$(Te); (d) Temperature dependence of in-plane resistivity normalized by the value at 300 K. (e) Complete phase diagram of FeSe$_{1-x}$Te$_x$ (0 $\leq x\leq$ 1) single crystals; (f) Enlarged view of (e) with temperature ranged from 0 to 30 K.}\label{}
	\end{figure*}
	
	\begin{figure*}\center
		\includegraphics[width=0.6\linewidth]{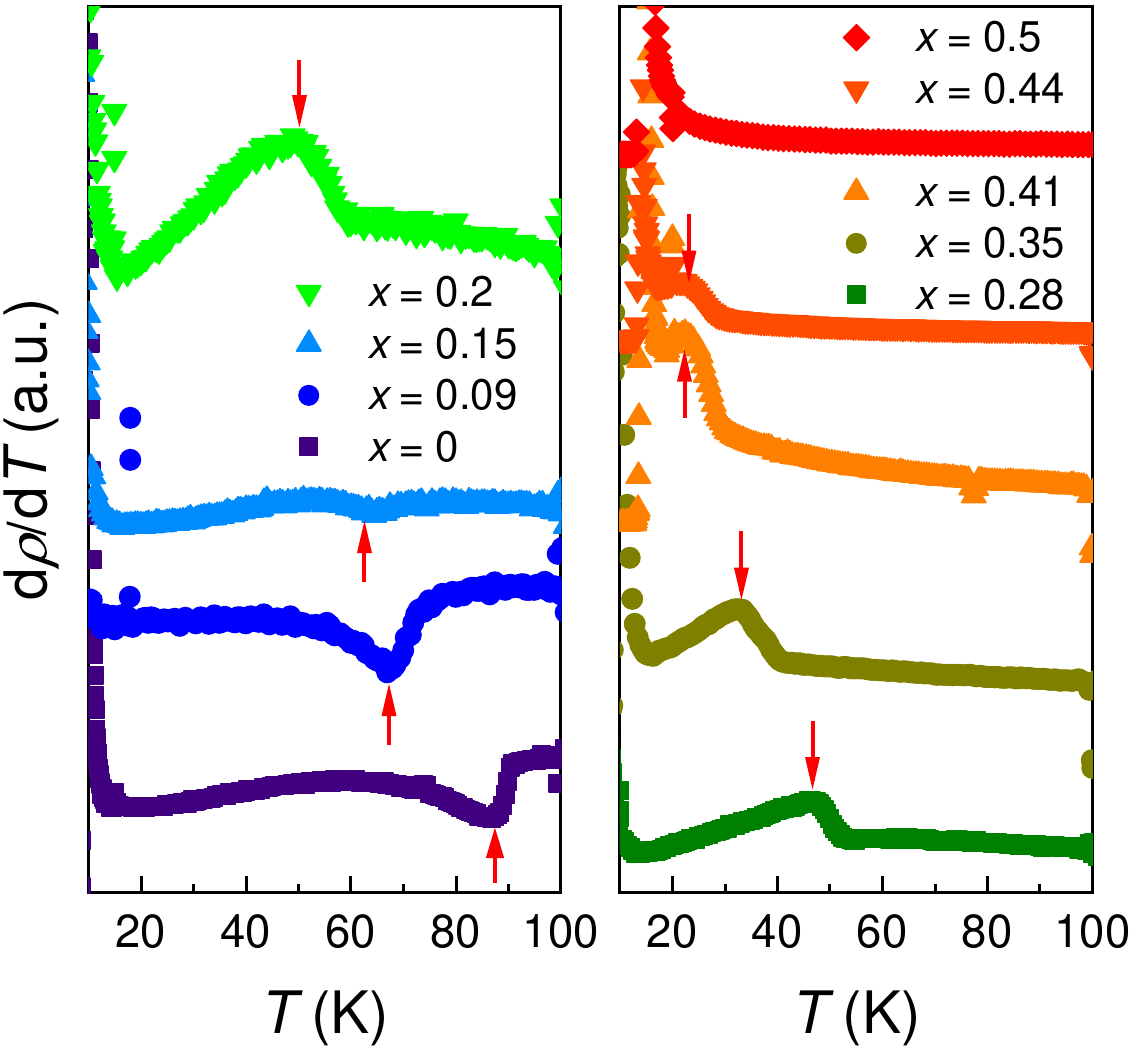}
		\caption{Temperature dependence of d$\rho$/d$T$ for 0 $\leq x\leq$ 0.5 in  FeSe$_{1-x}$Te$_x$ for determination of nematic (structural) transition temperature $T\rm_{s}$, marked by red arrow.}\label{}
	\end{figure*}

		\begin{figure*}\center
		\includegraphics[width=0.6\linewidth]{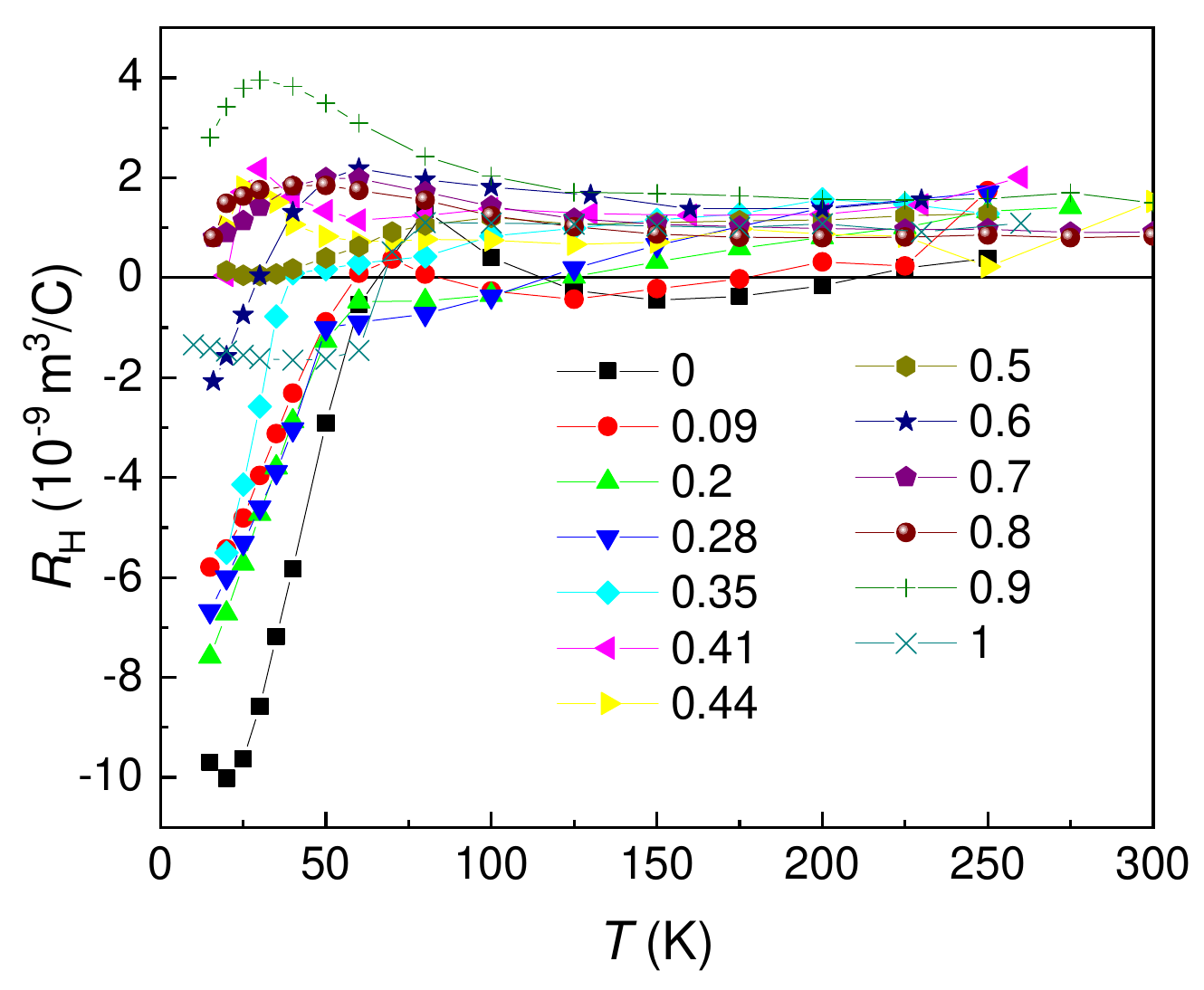}
		\caption{Temperature dependence of Hall coefficients $R\rm_{H}$ = $\rho_{xy}/\mu_0 H$ of all the FeSe$_{1-x}$Te$_x$ single crystals. In the case where exhibits nonlinear behavior,  $R\rm_{H}$ was determined from the linear region at small fields.}\label{}
	\end{figure*}
	
		\begin{figure*}\center
		\includegraphics[width=0.6\linewidth]{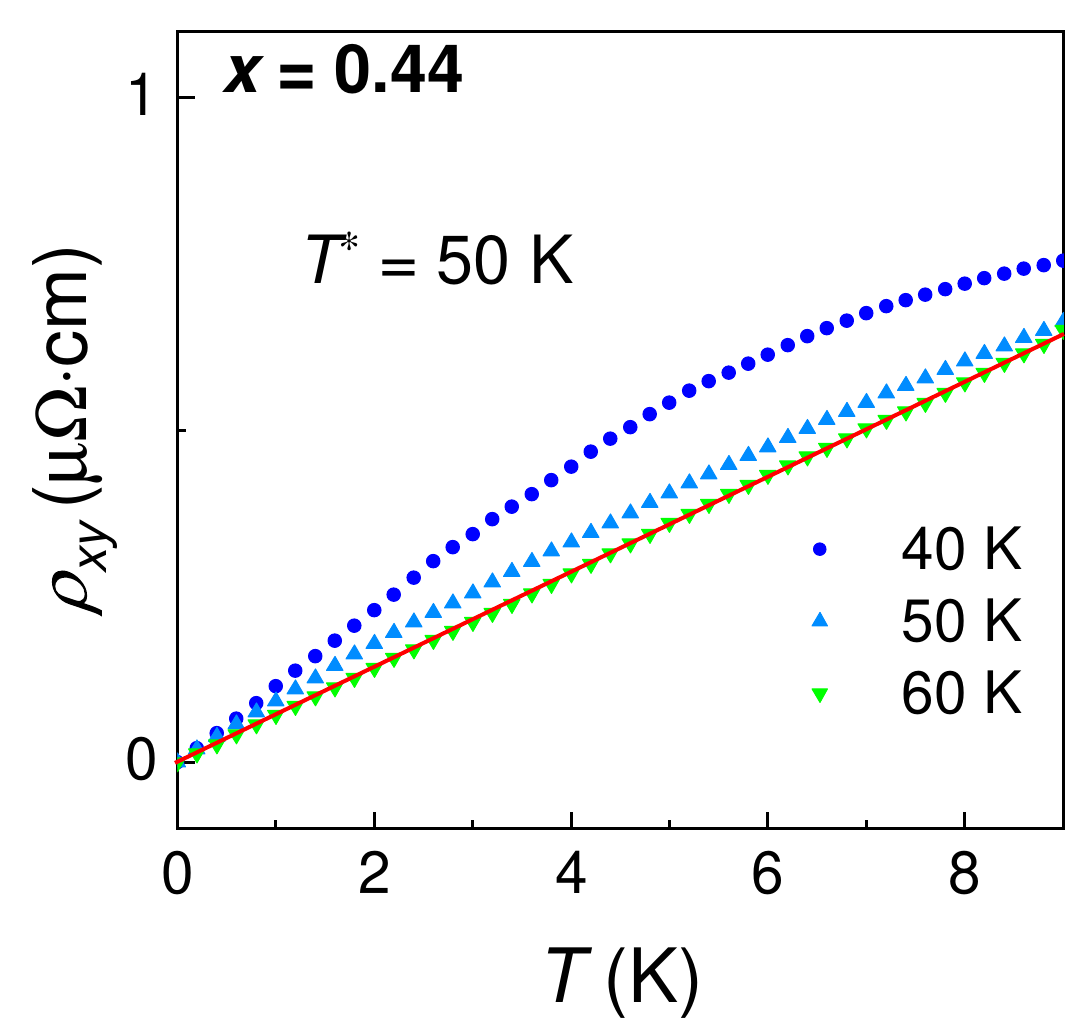}
		\caption{The temperature where the nonlinear Hall resistivity is located before the transition to linear is selected as the characteristic temperature $T^{*}$, taking FeSe$_{1-x}$Te$_x$ ($x$ = 0.44) as an example. $T^{*}$ is the critical temperature below which the Dirac states appear.}\label{}
	\end{figure*}
	
		\begin{figure*}\center
		\includegraphics[width=0.8\linewidth]{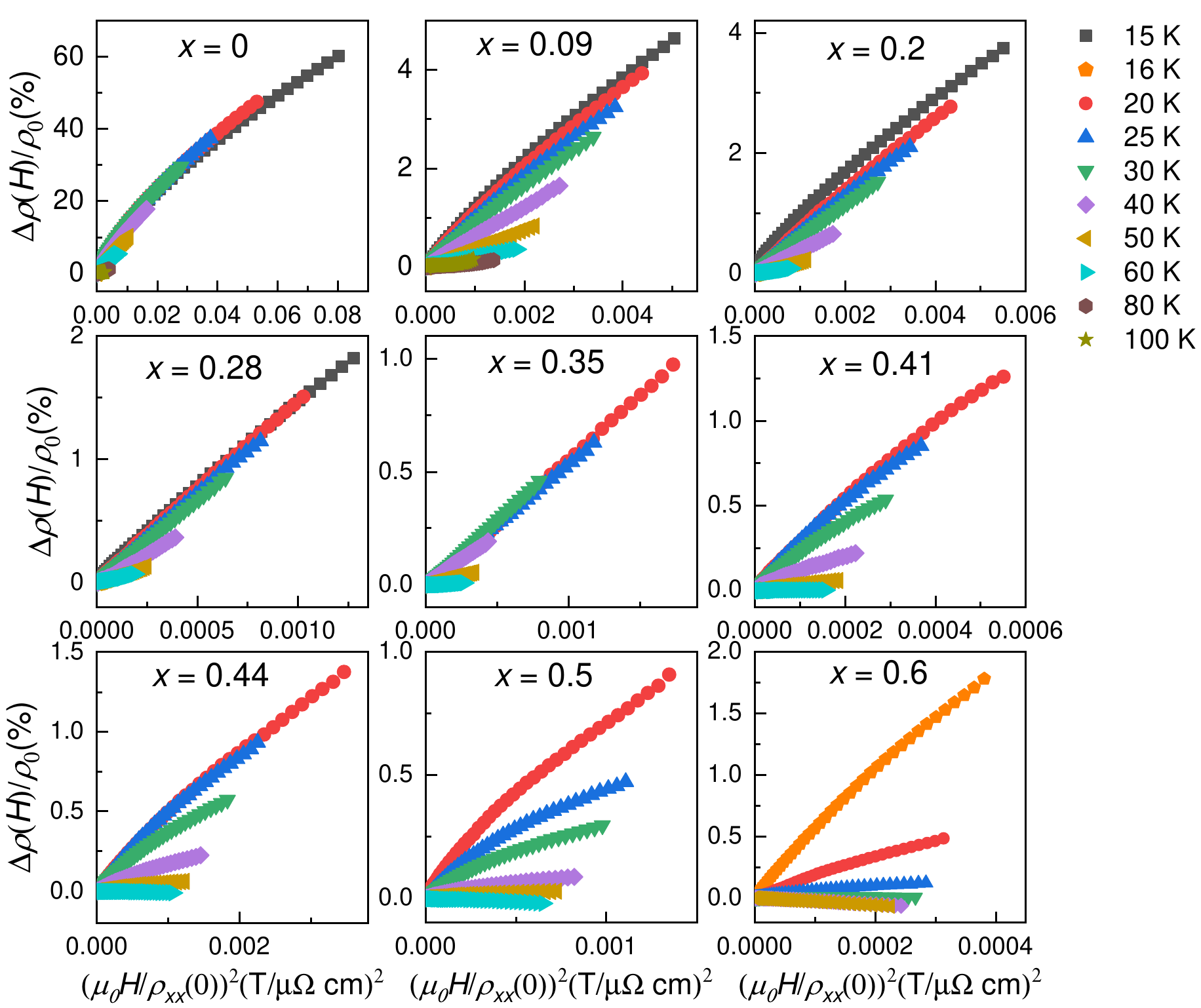}
		\caption{MR plotted as a function of ($\mu_0H/\rho_{xx}(0))^2$ at different temperatures for FeSe$_{1-x}$Te$_x$ (0 $\leq x\leq$ 0.6).}\label{}
	\end{figure*}
		
		\begin{figure*}\center
		\includegraphics[width=0.8\linewidth]{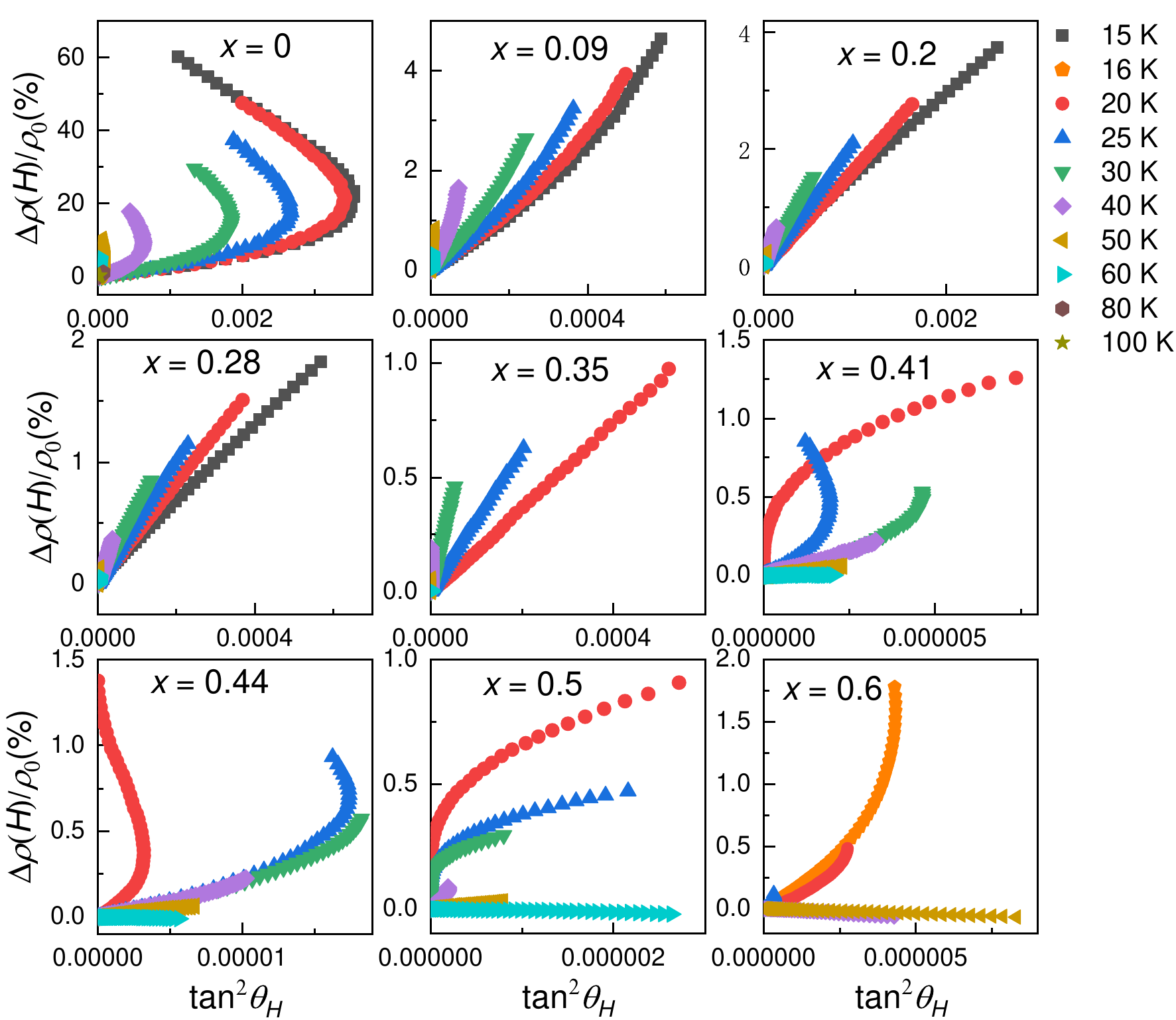}
		\caption{MR plotted as a function of $\rm{tan}^2\theta\rm_{H}$ at different temperatures for FeSe$_{1-x}$Te$_x$ (0 $\leq x\leq$ 0.6).}\label{}
	\end{figure*}
		
		\begin{figure*}\center
		\includegraphics[width=1\linewidth]{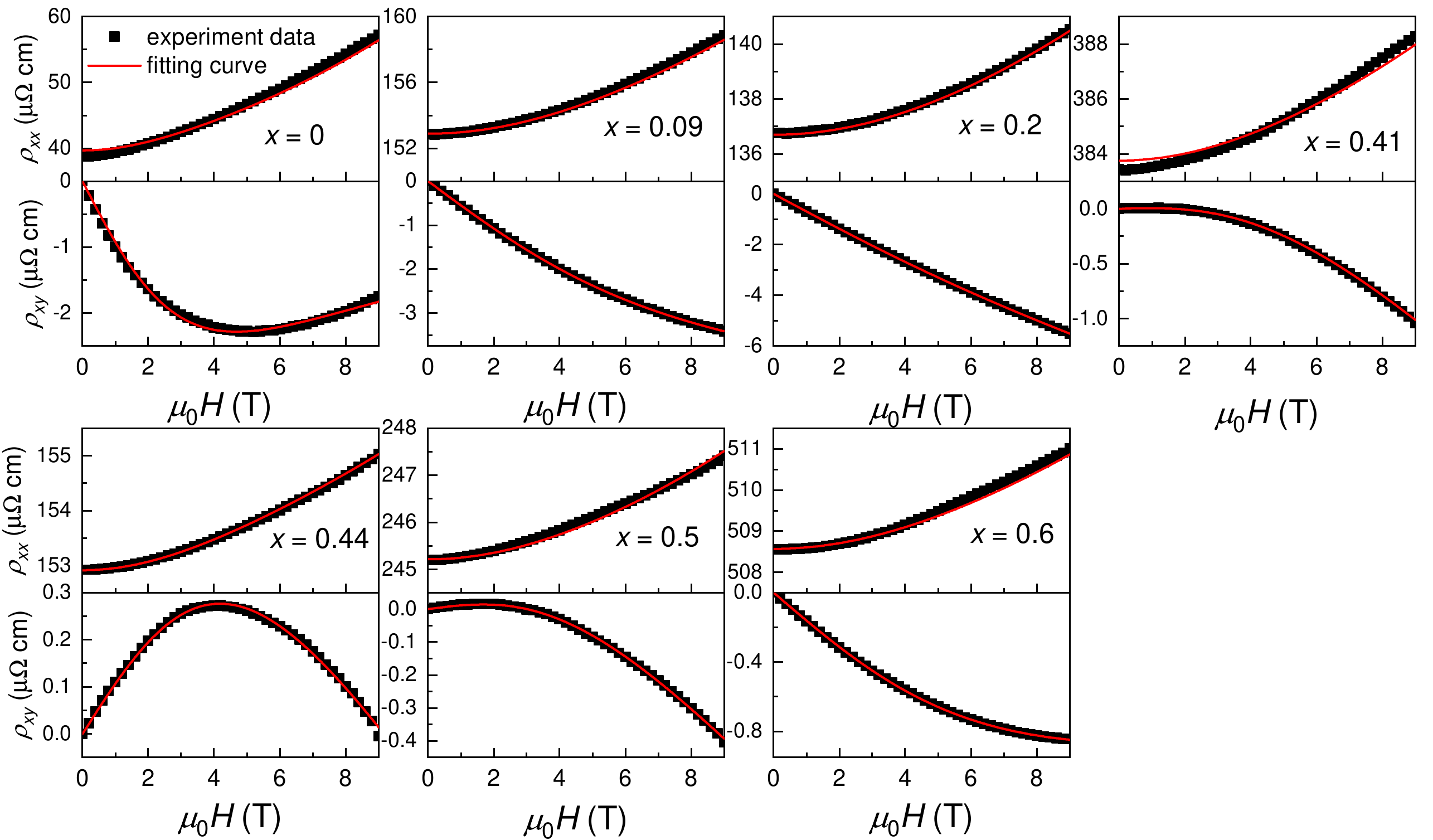}
		\caption{Simultaneous fitting of the MR and Hall curves at 20 K using a constrained compensated three-band model.}\label{}
	\end{figure*}

	\begin{table*}
		\caption{\label{tab:table1}Parameters extracted from magnetotransport measurements by three band model.}
		\begin{ruledtabular}
			\begin{tabular}{cccccccc}
				&$x$ = 0&0.09&0.2&0.41&0.44&0.5&0.6\\ \hline
				$n_h (10^{18} \rm{cm}^{-3})$&119.0&113.0&139.9&129.9	&197.8	&131.3	&97.6 \\
				$\mu_h (\rm{cm^{2}V^{-1}s^{-1}})$&612.6&	172.5&	143.8&	56.8&	99.7&	95.1&	62.7\\
				$n_e (10^{18} \rm{cm}^{-3})$&	111.2&	111.9&	139.2&	130.0&	198.1&	131.3&	97.5\\
				$\mu_e (\rm{cm^{2}V^{-1}s^{-1}})$&	589.8&	182.6&	178.6&	62.5&	104.5&	98.6&	62.9\\
				$n_2 (10^{18} \rm{cm}^{-3})$&		8.06&	1.04&	0.70&	0.07&	0.27&	0.06&	0.04\\
				$\mu_2 (\rm{cm^{2}V^{-1}s^{-1}})$&		2344.4&	1035.7&	1045.0&	1131.3&	1346.1&	1294.2&	1014.2\\
			\end{tabular}
		\end{ruledtabular}
	\end{table*}
	\clearpage
	
	\noindent
	\hangafter=1 [1]  N. Luo and G. H. Miley. Kohler’s rule and relaxation rates in high-$T\rm_{c}$ superconductors, \textit{Physica C: Superconductivity} 371:259, 2002.
	\setlength{\hangindent}{1.4em}
	
	\setlength{\hangindent}{1.4em} \noindent[2]  	M. Kohler. Zur magnetischen widerstandsänderung reiner metalle, \textit{Annalen Der Physik} 424:211, 1938.
	
	\setlength{\hangindent}{1.4em} \noindent[3]	J. M. Harris, Y. F. Yan, P. Matl, N. P. Ong, P. W. Anderson, T. Kimura, and K. Kitazawa, Violation of Kohler’s rule in the normal-state magnetoresistance of $\rm{YBa_2Cu_3O_{7-\delta}}$ and $\rm{La_2Sr_\textit{x}CuO_4}$, \textit{Phys. Rev. Lett.} 75:1391, 1995.
	
	\setlength{\hangindent}{1.4em} \noindent[4]	S. Kasahara, T. Shibauchi, K. Hashimoto, K. Ikada, S. Tonegawa, R. Okazaki, H. Shishido, H. Ikeda, H. Takeya, K. Hirata, T. Terashima and Y. Matsuda. Evolution from non-Fermi- to Fermi-liquid transport via isovalent doping in $\rm{BaFe_2(As_{1-\textit{x}}P_\textit{x})_2}$ superconductors, \textit{Phys. Rev. B} 81:184519, 2010.
	
	\setlength{\hangindent}{1.4em} \noindent[5]	M. J. Eom, S. W. Na, C. Hoch, R. K. Kremer, and J. S. Kim, Evolution of transport properties of BaFe$_{2-x}$Ru$_x$As$_2$ in a wide range of isovalent Ru substitution, \textit{Phys. Rev. B} 85:024536, 2012.
	
	\setlength{\hangindent}{1.4em} \noindent[6]	Xiangzhuo Xing, Yue Sun, Xiaolei Yi, Meng Li, Jiajia Feng, Yan Meng, Yufeng Zhang, Wenchong Li, Nan Zhou, Xiude He, Jun-Yi Ge, Wei Zhou, Tsuyoshi Tamegai, and Zhixiang Shi. Electronic transport properties and hydrostatic pressure effect of FeSe$_{0.67}$Te$_{0.33}$ single crystals free of phase separation, \textit{Supercond. Sci. Technol.}, 34:055006, 2021.
	
\end{document}